\documentclass{aa}
\usepackage{natbib}
\bibpunct{(}{)}{;}{a}{}{,} 
\usepackage{graphicx}
\usepackage{txfonts}
\usepackage{longtable}
\usepackage{amsfonts}
\newcommand \kms {\mbox{km~s$^{-1}$}}
\newcommand \degree {\mbox{$^\circ$}}
\begin{document}
	\title{TANGO I: \\ISM in nearby radio galaxies. Molecular gas}

   \author{B. Oca\~na Flaquer
          \inst{1}
          \and
          S. Leon \inst{2,1}
	   \and
	   F. Combes \inst{3}
	   \and 
	   J. Lim \inst{4,5}
          }

   \offprints{B. Oca\~na Flaquer}

   \institute{Instituto de Radio Astronom\'ia Milim\'etrica (IRAM),
	      Av. Divina Pastora, 7, N\'ucleo Central, 18012 Granada, Spain\\
              \email{ocana@iram.es}
	  \and 
	      Joint Alma Observatory, Av. El Golf 40, Piso 18, Las Condes, Santiago, Chile
	\and
             Observatoire de Paris, LERMA, 61 Av. de l'Observatoire, F-75014. Paris, France\\
             \email{francoise.combes@obspm.fr}
        \and
	       Institute of Astronomy and Astrophysics, Academia Sinica,128, Sec. 2, Yen Geo Yuan Road, Nankang, Taipei, Taiwan, R.O.C. 
		\email{lim@asiaa.sinica.edu.tw}
    \and
		Department of Physics, University of Hong Kong, Pokfulam Road, Hong Kong\\
			\email{jjlim@hku.hk}
             }

   \date{Received MONTH day, YEAR; accepted MONTH day, YEAR}

  \abstract
  {Powerful radio-AGN are hosted by massive elliptical galaxies which are usually very poor in molecular gas. Nevertheless gas is needed in the very center to feed the nuclear activity.}
   {We aim to study the molecular gas properties (mass, kinematics, distribution, origin) in such objects, and to compare them with results of other known samples.}
   {We have performed at the IRAM-30m telescope a survey of the CO(1-0) and  CO(2-1) emission in the most powerful radio galaxies of the Local Universe, selected only on the basis of their radio continuum fluxes.} 
   {The main result of our survey is the very low content in molecular gas of such galaxies compared to spiral or FIR-selected galaxies. The median value of the molecular gas mass, including detections and upper limits, is $2.2\times10^8~M_{\odot}$. If separated into FR-I and FR-II types, a difference in $H_2$ masses between them is found. The median value of FR-I galaxies is about $1.9\times10^8$ and higher for FR-II galaxies, about $4.5\times10^8~M_{\odot}$  but this is very probably entirely due to a Malmquist bias. Our results contrast with previous surveys, mainly selected through the FIR emission, implying larger observed masses of molecular gas. Moreover, the shape of CO spectra suggest the presence of a central molecular gas disk in 30\% of these radio galaxies, a lower rate than in other active galaxy samples. } 
{We found a low level of molecular gas in our sample of radio-selected AGNs, indicating that galaxies do not need much molecular gas to host an AGN. The presence of a molecular gas disk in some galaxies and the large range of molecular gas mass may indicate different origins of the gas that we can not discard at present: minor/major merger, stellar mass loss or accretion.} 
   \maketitle
%

\section{Introduction}
Radio galaxies have strong radio sources in the range of $10^{41}$ to $10^{46}$ erg/s, equivalent to $10^{34}-10^{39}$ Watts,  normally hosted by giant elliptical galaxies with visual luminosities of about 2.1$\times$10$^{10} h^{-2}$ L$_\odot$ (\cite{Kellerman88}, calculated with a Hubble constant of H=100 km.s$^{-1}$Mpc$^{-1}$). Although in general relatively deficient in cold gas, elliptical galaxies still contain some dense and cold Interstellar Medium (ISM) detected in CO(1-0) emission \citep[e.g.][]{Wiklind86}. Dust is also detected in the far-infrared (FIR) radiation which is assumed to be thermal and to originate from dust heated by young massive stars or by the active galactic nuclei (AGN) \citep{knapp89,Wiklind95b}. According to \cite{Kennicutt98a}, early-type galaxies show no independent evidence of high star formation rates (SFRs), suggesting that the older stars or the AGN are responsible for much of the FIR emission.\\
\cite{Wiklind95b} show that in elliptical galaxies the gas is unrelated to the stellar populations, and favor an external origin of the molecular gas.\\
It is widely believed that the AGN are powered by accretion of ISM onto a super massive black holes (SMBHs) \citep{Antonucci93}. According to \cite{deRuiter02} the presence of ISM in the circumnuclear regions of AGN is indeed inevitable and the large-scale dust/gas systems should be related to nuclear activity.\\
More recent studies by \cite{Hardcastle07} suggest that it is possible that all the apparently different accretion modes may be the result of a different source for the accreting gas. This hypothesis came  in \cite{Allen06} who showed that some low luminosity radio galaxies in the center of clusters could be powered by Bondi accretion, a spherical accretion of the hot, X-ray emitting medium. The supporting arguments about the nature of the accretion mode in low power radio sources were discussed as well  by \cite{Best06}.\\
For low-luminosity AGN, the stellar component might be sufficient to fuel the active nucleus, but for powerful radio galaxies large scale dynamical processes are required to transfer angular momentum of the interstellar gas across the disk radius \citep{Shlosman90,Combes02}.\\
Fueling is also dependent on the morphological type of the galaxy, since the radial gas flows are driven by non-axisymmetric instabilities, and the stability of the disk is essentially dependent on the bulge-to-disk ratio \citep{Mihos96,Mihos94}.  In early type galaxies, for example, once the accreted gas has settled in a symmetric disk, the large-mass black holes (BH) are starving, unless perturbed by an external tide \citep[e.g.][]{Combes02}.\\
In order to better understand the role of the Interstellar Medium (ISM) in the radio galaxies we have built two samples, at low and medium redshift, of  radio galaxies selected only on the basis of their radio continuum emission. \\
This is the first paper of the Thorough ANalysis of radio-Galaxies Observation (TANGO) project of a series of papers that will cover a wide range in wavelength (optical, radio, IR, X-Ray). \\
An observational criterion of the TANGO samples at   low and medium redshift is that most of the objects are visible from the main (radio) telescopes  of the the Northern and Southern hemisphere, such as  ALMA, IRAM telescopes, VLA, VLT, GTC among others.\\
The primary issue addressed in the present paper is the study of the molecular gas through the CO emission in the low redshift  TANGO  sample of radio galaxies (z$\leq$0.1). We want to test, in particular, whether samples selected on the FIR emission, e.g. \cite{Evans05}, and radio continuum-selected samples behave in similar ways.\\
This paper is structured as follows: we describe the sample of radio galaxies, their observations and their data reduction process in $\S$ \ref{obs}. The molecular gas properties are described in $\S$ \ref{mh2ch}, while the dust properties are described in $\S$ \ref{dust} and the gas and dust relations are discussed in $\S$ \ref{firco}. Section $\S$ \ref{sec:comp} is dedicated to the comparison of the result of this work with similar samples. Finally the discussion and the conclusions are drawn in $\S$ \ref{discussion} and $\S$ \ref{conclusion} respectively. Properties of individual galaxies are presented in the Appendix \ref{ig}, the beam/source coupling is explained in details in the  Appendix \ref{coupling} and the Appendix \ref{3c31} is dedicated to the detail analysis of one of the galaxies in TANGO, 3CR 31.\\
%
\section{Observation and data reduction} \label{obs}
%
\subsection{Sample}
As previously mentioned in the introduction, the  galaxies selected in the low and medium redshift TANGO sample  have one main  criterium of selection: a high radio continuum power. Thus in the low redshift sample studied in this paper, all the radio galaxies have a radio power at 1.4 GHz larger than $10^{22.5}$  W.Hz$^{-1}$ with a median of $10^{24.4}$  W.Hz$^{-1}$.  This makes the main  difference with the  previous samples, chosen e.g. by their FIR flux \citep[e.g.,][]{Evans05}. \\
This work studies a total of  52 nearby radio galaxies, most of them from the Third Cambridge Catalog \citep[3CR -- ][]{Edge59, Bennett62}, New General Catalog \citep[NGC -- ][]{Dreyer88}, and the Second Bologna Catalog of Radio Sources \citep[B2 -- ][]{Colla70, Colla72, Colla73, Fanti74}. There is one galaxy from the Ohio State University Radio survey Catalog  \citep[OQ --][]{Scheer67} and one galaxy from the Uppsala General Catalogue of Galaxies \citep[UGC -- ][]{Nilson73}. The details of the catalogs for the sample are summarized in Table \ref{catalog}.
\begin{table}[!h]
\centering
\begin{tabular}{|c|c|c|}
\hline 
\textbf{Catalog Name}&\textbf{$\#$ of Galaxies}&\textbf{$\%$ of detection $^1$ }\\
\hline 
3CR & 24 & 58$\%$ \\
NGC & 13 & 62$\%$ \\
B2 & 13 & 31$\%$\\
UGC & 1 & 0.0$\%$\\
OQ & 1 & 100.0$\%$\\
\hline 
\textbf{Total} & 52 & 52$\%$ \\
\hline
\end{tabular}
\caption{The catalogs are listed in order of importance, some galaxies are in more than one catalog with different names. In this article the names are chosen first in favor of the 3CR catalog and last for the OQ catalog.
 $^1$ This $\%$ is within the catalog, meaning the percentage of galaxies detected in each catalog.}\label{catalog}
\end{table}
\subsection{Observations}
This work is based on a survey of the  $^{12}$CO(1-0) and $^{12}$CO(2-1) transitions, simultaneously observed with the IRAM 30 m telescope, at Pico Veleta - Spain, reaching an rms temperature of about 0.6 mK with a velocity resolution between 20 km/s and 52 km/s. \\
The transitions were observed at 115.27 GHz and 230.54 GHz (where the beam of the IRAM-30m telescope is about 22 and 11 arcsec), redshifted at the velocity of the galaxies, for the $^{12}$CO(1-0) and $^{12}$CO(2-1) respectively with the A and B receivers. The filter backends were configured into two units of 512 x 1 MHz channels and two units of 512 x 4 MHz, one for each receiver. During the observations, the pointing and focus were monitored by observing planets and standard continuum sources every hour with an accurancy of about 3 arcseconds.\\
From the total 52 galaxies in the sample, all of them were observed in the CO(1-0) line and 43 galaxies have simultaneously been observed in CO(2-1) line, when the weather permitted. About 58\% of galaxies (N=30) in the sample have been detected (clearly or tentatively) in one or both lines and 38\% of the galaxies (N=20) have been clearly detected, some of them only in CO(1-0), some only in CO(2-1) and some in both CO transitions. A total of 7 galaxies have been detected in both lines, these galaxies are 3CR 31, 3CR 264, NGC 326, NGC 4278, NGC 7052, B2 0116+31 and B2 0648+27. These 7 galaxies have been used to study the CO(2-1)-to-CO(1-0) line ratios that will be discussed in section $\S$ \ref{lineratios} in more detail. \\
There were 9 galaxies for which we only have data in the 3 mm wavelength range (3CR 236, 3CR 305, 3CR 321, 3CR 327, 3CR 403, 3CR 433, NGC 6251, B2 0836+29 and OQ 208) where 6 of them have shown a clear detection (3CR 305, 3CR 321, 3CR 327, 3CR 403, B2 0836+29 and OQ208).\\ 
Table \ref{dates} is a journal of the observations which shows the dates on which each detected galaxy was observed. \\
This project started in 1999 and the last observations are from the year 2007, some of the galaxies have been observed several times throughout these 8 year period to confirm some possible detections. Part of this work has already been published by \cite{Lim2000} with a detailed study of the galaxies 3C31 and 3C264.\\
We observed as well the CO(1-0) emission in the center of 3C31 using the IRAM Plateau de Bure Interferometer (PdBI). We used several tracks in the B configuration. The velocity resolution was set to 20 \kms and the CLEANed maps have been restored with a synthesized beam of $2.\arcsec 2 \times 1.\arcsec 2$ (PA=47$\degree$). See Appendix \ref{3c31} for a clear analysis.\\
\begin{table} 
\centering 
\begin{tabular}{|l | l || l|l|}        
\hline
\textbf{Galaxy} & Observation Dates & \textbf{Galaxy} & Observation Dates \\ \hline
$3C31$	& 1999 Dec	& $NGC315$	& 2002 Jun\\
	& 2000 Jan	& 		& 2006 Oct		\\
$3C66BB$& 2000 Jan	& 		& 2007 Apr	\\
	& 2003 May	& $NGC326$	& 2006 Oct	\\
$3C83.1$& 2000 Jan	& $NGC541$	& 2002 Jun	\\
$3C88$	& 1999 Dec	& $NGC708$	& 2006 Oct\\
	& 2000 Jan	& $NGC3801$	& 2002 Jun\\
$3C129$	& 2000 Jan	& $NGC4278$	& 2007 Jan	\\
$3C264$	& 2000 Jan	& 		& 2007 Apr\\
$3C272.1$&2000 Jan	& $NGC5127$	& 2002 Jun\\
$3C274$	& 2000 Jan 	& $NGC7052$	& 2002 Jun\\
$3C305$	& 2004 Feb	& 		& 2006 Jul	\\
	& 2004 Mar	& $B2~0116+31$	& 2003 Dec\\
$3C321$	& 2004 Mar	& 		& 2006 Jun	\\
$3C327$	& 2004 Mar	& 		& 2006 Oct	\\
$3C353$	& 2000 Jan	& $B2~0648+27$	& 2004 Feb\\
$3C386$	& 1999 Dec	& 		& 2004 Mar	\\
$3C403$	& 2003 Dec	& 		& 2006 Oct\\
	& 2004 Feb	& $B2~0836+29B$	& 2004 Mar\\
$3C442$	& 2000 Jan	& $B2~0924+30$	& 2007 Apr	\\
$3C449$	& 1999 Dec	& $B2~1347+28$	& 2007 Jan\\
	& 2000 Jan	& 		& 2007 Apr\\
	&		& $OQ208$	& 2004 Mar\\
\hline
\end{tabular}
\caption{\newline Observations journal of the 30 detected galaxies in the sample.}   \label{dates}          
\end{table}
\subsection{Data reduction}
For all the data reduction, the Continuum and Line Analysis Single-dish Software (CLASS) was used. All spectra were individually checked, the bad ones were removed, and the remaining spectra of the same frequency were added.  The baseline was removed after selecting a window. This window was taken to be the velocity width for the detected galaxies, and a line-width value of $\sim$300 km/s was applied to compute the  upper limits. Each spectra was smoothed to a velocity resolution between 20 and 50 km/s. The integrated intensity, the velocity width, position and peak temperature of the galaxies were determined by fitting a Gaussian. The result spectra of each detected galaxy can be found in the online material.\\ 
We use the baseline of the spectra to calculate the continuum flux at 3 and 1 mm. The continuum level and the detections were given in antenna temperature, $T_a^*$, and then converted to main beam temperature, $T_{MB}$. The conversion factor used was the one given by \cite{Rohlfs04}, where we divided the forward efficiency (which is a model function to the measure antenna temperature and the copper load temperature) by the beam efficiency (which is the percentage of all power received that enters the main beam), $F_{eff}/B_{eff}$. Applying this for each frequency in the case of the IRAM-30m telescope, the relationship is 1.27 at 3mm and 1.75 at 1mm.\\
To convert the data from temperature (K) to flux (Jy), the point source sensitivity measurement (S/$T_A^*$) is 6.3 for 3mm and 8 for 1mm wavelength. The results for these conversions are listed in Table \ref{fl30m}. For more details, on the conversion see \cite{Kramer97}. From the observations, in summary, 90\% out of the 52 galaxies were detected in the continuum at 3mm and 65\% out of the 43 galaxies were detected in the continuum at 1mm.
\begin{table}[!ht] 
\centering 
\begin{tabular}{l | c c c c}        
\hline
\textbf{Galaxy} &$S_{3mm}$	& $\delta S_{3mm}$	&$S_{1mm}$	&$\delta S_{1mm}$ \\
		& (mJy)		& (mJy)	&(mJy) &	(mJy)	\\ \hline
$3C31$	& 79.07	& 1.13	& 74.39	& 2.78	\\
$3C40$	& 96.91	& 0.9	& 20.11	& 1.8 	\\
$3C66B$	& 110.12& 0.43	& 73.08	& 1.39	\\
$3C83.1$& 33.45	& 0.69	&  ...	& ... 	\\
$3C88$	& 121.84& 0.43	& 87.05	& 1.3	\\
$3C98$	& 18.09	& 0.52	& 6.18	& 1.54	\\
$3C129$	& 28.53	& 0.65	& 0.91	& 1.27	\\
$3C236$	& 225.06& 0.57	&  X	&  X	\\
$3C264$	& 174.64& 0.5	& 87.35	& 1.13	\\
$3C270$	& 326.34& 0.82	& 204.71& 1.57	\\
$3C272.1$&138.29& 1.2	& 73.86	& 6.79	\\
$3C274$	&2347.13& 1.83	&1312.31& 2.78 	\\
$3C296$	& 99.23	& 0.6	& 54.55	& 1.22	\\
$3C305$	& 21.23	& 4.28	&  X	&  X	\\
$3C321$	& 67.33	& 0.52	&  X	&  X	\\
$3C327$	& 105.81& 0.69	&  X	&  X	\\
$3C353$	& ...	& ...	& 3.46	& 1.35	\\
$3C386$	& 50.21	& 0.88	& 105.01& 2.26	\\
$3C402$	& 332.57& 39.27	& ...	& ...	\\
$3C403$	& 20.35	& 0.48	&  X	&  X	\\
$3C433$	& 91.29	& 0.76	&  X	&  X	\\
$3C442$	& 9.62	& 0.43	& 75.51	& 9.99	\\
$3C449$	& 25.82	& 0.5	& 15.83	& 1.65	\\
$3C465$	& 86.06	& 0.56	& 27.32	& 1.31	\\
$NGC315$& 353.56& 0.67	& 62.12	& 1.39	\\
$NGC326$& 63.47	& 0.63	& ...	& ...	\\
$NGC541$& 4.73	& 2.02	& 27.23	& 2.60 	\\
$NGC708$& 8.11	& 0.98	& 148.94& 3.83	\\
$NGC2484$	& 110.63	& 0.69	& 11.92	& 1.65 	\\
$NGC2892$	& 190.05	& 0.87	& 272.39& 2.23 	\\
$NGC3801$	& 1181.72	& 210	& ...	& ... 	\\
$NGC4278$	& 215.9		& 1.08	& 194.34& 1.28 	\\
$NGC5127$	& 25.62		& 1.13	& ...	& ...	\\	
$NGC5141$	& 63.50		& 1.01  & ...	& ... 	\\ 
$NGC5490$	& 27.72		& 1.44	& 88.61	& 5.46 	\\
$NGC6251$	& 367.88	& 0.71	& X	& X 	\\
$NGC7052$	& 85.51		& 12.13	& 18.77	& 1.49 	\\
$B2~0034+25$	& 72.97		& 0.68	& 190.81& 1.82	\\
$B2~0116+31$	& 164.53	& 1.22	& ...	& ...	\\
$B2~0648+27$	& 152.99	& 0.37	& 172.99& 2.09	\\
$B2~0836+29B$	& ...		& ...	& X	& X	\\
$B2~0915+32B$	& 11.39		& 1.29	& 24.41	& 2.17	\\
$B2~0924+30$	& 9.44		& 1.13	& ...	& ...	\\ 
$B2~1101+38$	& 359.89	& 0.97	& 0.165	& 0.002	\\
$B2~1347+28$	& 8.96		& 0.51	& ...	& ...	\\
$B2~1357+28$	& ...		& ...	& ...	& ...	\\
$B2~1447+27$	& 13.94		& 0.53	& 52.41	& 0.81	\\
$B2~1512+30$	& ...		& ...	& ...	& ...	\\
$B2~1525+29$	& 14.48		& 1.07	& 0.13	& 0.003	\\
$B2~1553+24$	& 35.28		& 1	& ...	& ...	\\
$UGC7115$	& 30.57		& 1.03	& ...	& ...	\\
$OQ208$		& 40.43		& 16.67	& X	& X	\\ 
\hline
\end{tabular}
\caption{IRAM 30m fluxes at 3 mm and 1 mm.  \textit{ X means that there was no data, and "...'' refers to undetected flux.}} 
\label{fl30m}
\end{table}
\subsection{IRAS Data} \label{iras}
We use the IRAS data to get the dust properties both for the warm component ($\lambda \sim 60 \mu m$) associated with young star forming regions and/or an AGN, and the cooler components ($\lambda \geq 100 \mu m$) associated with more extended dust heated by the interstellar radiation field \citep{Kennicutt98a}. We complemented the 30m data with the IRAS fluxes taken from the Nasa Extra-galactic Database (NED) for the 60$\mu$m and 100$\mu$m wavelengths. From this sample, a total of 38 galaxies were detected by IRAS. 18 of them are considered a clear detection; 15 galaxies are upper limits for both 60$\mu$m and 100$\mu$m wavelength and 5 galaxies are upper limit only at 100$\mu$m and detected at 60$\mu$m. For that study we will use hereafter the 18 galaxies that were clearly detected in both wavelengths, the galaxies and their fluxes are listed on Table \ref{iras-data}.
\begin{table}[!ht]
	\centering
	\begin{tabular}{|l||l|l|}
		\hline
		\textbf{Galaxy Name} & \textbf{$f_{60}~\mu$m} & \textbf{$f_{100}~\mu$m}\\
		                     &     (mJy)                &   (mJy)                 \\
		\hline
		3CR 31      & 435  $\pm$ 65.3 & 1675 $\pm$ 251  \\
		3CR 88      & 180  $\pm$ 27	  & 816  $\pm$ 122  \\
		3CR 272.1   & 556  $\pm$ 83.4 & 1024 $\pm$ 154  \\
		3CR 274     & 546  $\pm$ 81.9 & 559  $\pm$ 83.9 \\
		3CR 305     & 298  $\pm$ 44.7 & 450  $\pm$ 67.5 \\
		3CR 321     & 1067 $\pm$ 160  & 961  $\pm$ 144  \\
		3CR 327     & 670  $\pm$ 101  & 371	 $\pm$ 55.7 \\
		3CR 402     & 257  $\pm$ 38.6 & 1052 $\pm$ 158  \\
		NGC 315     & 368  $\pm$ 55.2 & 460  $\pm$ 69   \\
		NGC 708     & 200  $\pm$ 34   & 660  $\pm$ 157  \\
		NGC 4278    & 580  $\pm$ 53   & 1860 $\pm$ 60   \\
		NGC 6251    & 188  $\pm$ 28.2 & 600  $\pm$ 90   \\
		NGC 7052    & 524  $\pm$ 78.6 & 1150 $\pm$ 173  \\
		B2 0116+31  & 150  $\pm$ 22.5 & 524  $\pm$ 78.6 \\
		B2 0648+27  & 2633 $\pm$ 395  & 1529 $\pm$ 229  \\
		B2 0836+29B & 472  $\pm$ 70.8 & 595  $\pm$ 89.2 \\
		B2 1101+38  & 181  $\pm$ 22   & 361  $\pm$ 68   \\
		OQ 208      & 753  $\pm$ 113  & 1029 $\pm$ 154  \\
		\hline
	\end{tabular}
	\caption{IRAS data for the galaxies that have been detected at 60 $\mu$m and at 100 $\mu$m.}\label{iras-data}
\end{table}

%
\section{Molecular Gas} \label{mh2ch}
\subsection{Molecular gas mass} \label{mmass}
\begin{figure} 
\centering
\includegraphics[scale=0.45]{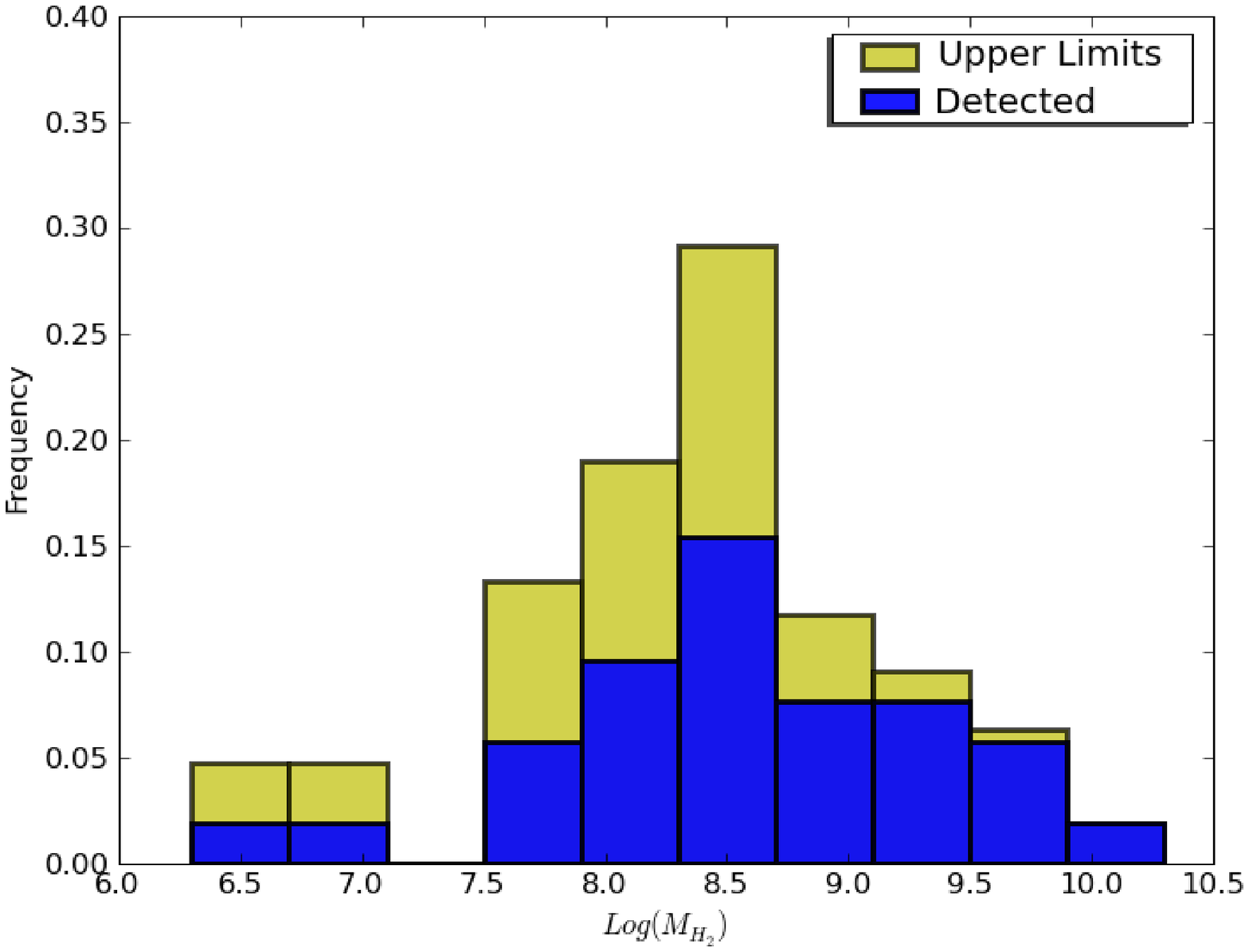}
\caption{This figure represents the $M_{H_2}$ histogram of the sample. The lighter color represents all the galaxies in the sample. Using ASURV we were able to calculate the mass of the sample using both, upper limits and  detected galaxies. Plotted at the bottom, with the darker color, are the detected galaxies only.}
\label{mh2histo}
\end{figure}

\begin{figure} 
\centering
\includegraphics[scale=0.45]{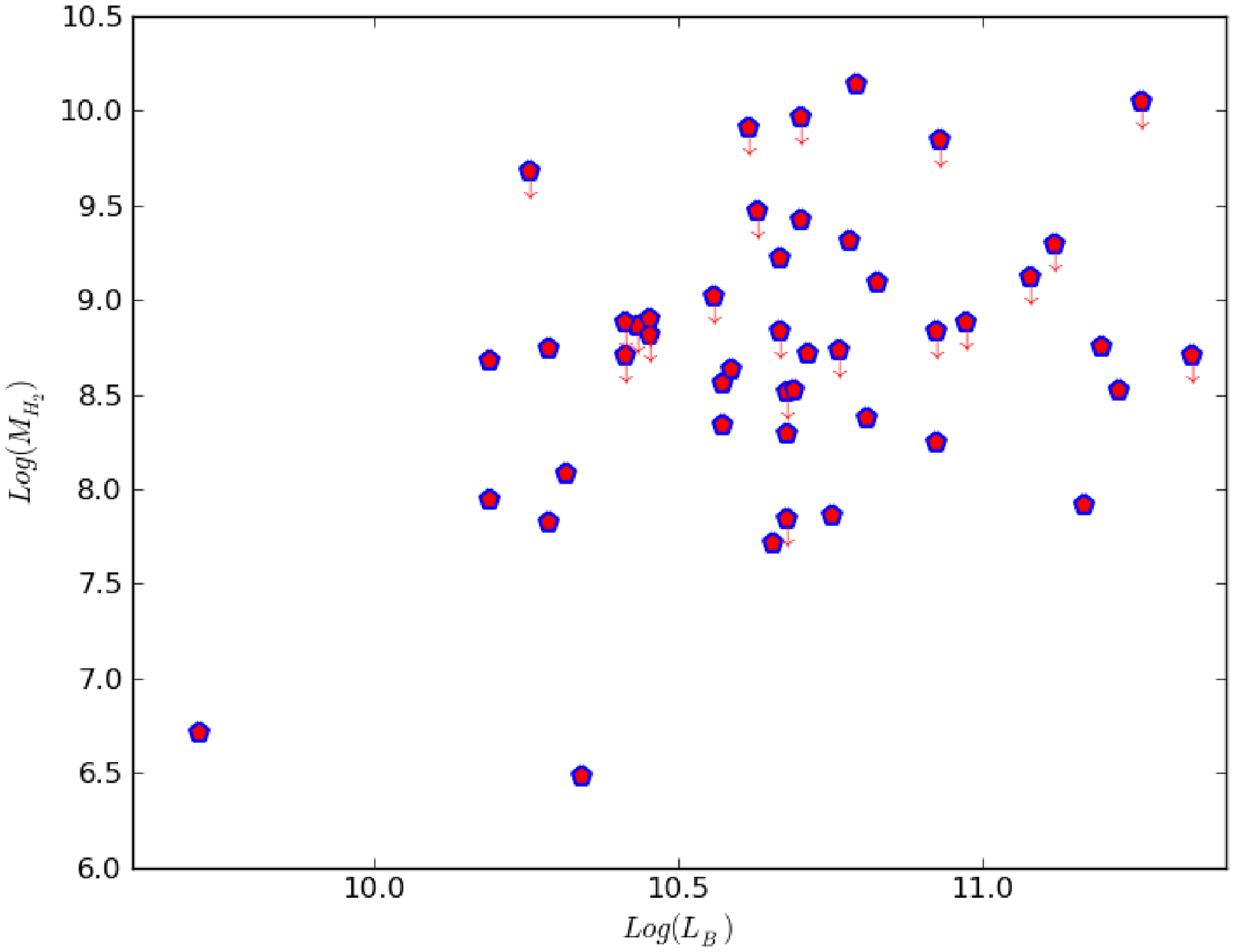}
\caption{Molecular gas mass versus blue luminosity with the upper limits for the molecular gas mass indicated.}\label{fig:lbmh2}
\end{figure}

\begin{figure} 
\centering
\includegraphics[scale=0.45]{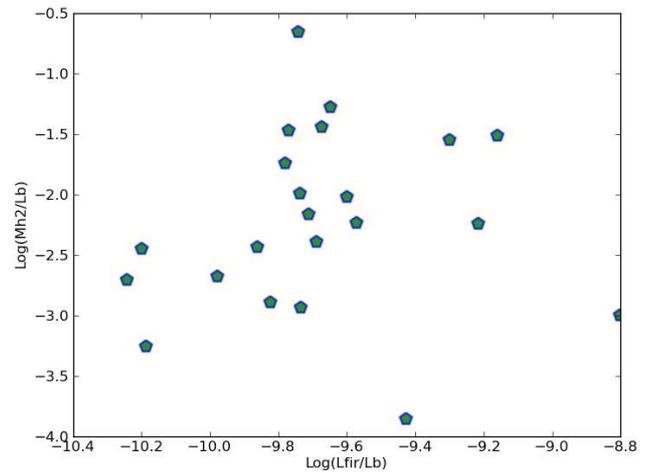}
\caption{Molecular gas mass versus FIR luminosity, both normalized by the blue Luminosity, using only detected values.}\label{fig:lbmh2lfirlb}
\end{figure}

%
In Table \ref{mol} we present the results of the observations, where $I_{CO(1-0)}$ and $I_{CO(2-1)}$ are the integrated intensities ($\int T_{MB}d\upsilon$) for the $^{12}$CO(1-0) and  $^{12}$CO(2-1) line respectively,  $\delta I_{CO(1-0)}$ and $\delta I_{CO(2-1)}$ are the standard error on $I_{CO(1-0)}$ and $I_{CO(2-1)}$, V$_{width~ CO(1-0)}$ and V$_{width~ CO(2-1)}$ are the velocity width of CO(1-0) and CO(2-1) respectively; and finally, $M_{H_2}$ is the calculated molecular gas mass for each galaxy in the sample.\\ 
We follow \cite{Gordon92} to derive H$_2$ masses from the $^{12}$CO(1-0) line observations. Our temperature unit is expressed in $T_A^*$ antenna temperature scale, which is corrected for atmospheric attenuation and rear side-lobes. The radiation temperature $T_R$ of the extragalactic source is then:
\begin{equation} \label{tr}
T_R = \frac{4}{\pi}(\frac{\lambda}{D})^2\frac{K}{\eta_A}\frac{T_A}{\Omega_S}
\end{equation}
where $\lambda$ is the observed wavelength (2.6 mm), $D$ is the IRAM radio telescope diameter (30 m), $K$ is the correction factor for the coupling of the sources with the beam, $\eta_A$ is the aperture efficiency (0.55) at 115 GHz, $T_A$ is an antenna temperature which is $F_{eff}T_A^*$ in the IRAM convention, explicitly $T_A=0.92T_A^*$, and $\Omega_S$ is the source solid angle.\\
Without taking into account cosmological correction, because of the low redshift and using the standard CO-to-H$_2$ conversion factor, $2.3\times10^{20} cm^{-2} (K ~ km ~s^{-1})^{-1}$ suggested by \cite{Strong88}, the  column density of molecular hydrogen can be written as:
\begin{equation} \label{nh2}
N(H_2) = 2.3 \times 10^{20} \int_{line} T_Rd\nu (mol.cm^{-2})
\end{equation}
Where $d\nu$ is the velocity interval. In Eq. \ref{tr} $K \equiv \Omega_s/\Omega_{\Sigma}$ is the factor which corrects the measured antenna temperature for the weighting of the source distribution by the large antenna beam in case of a smaller source. We have defined the source solid angle as:
\begin{equation} \label{oms}
\Omega_S \equiv \int_{source} \phi(\theta,\psi)d\Omega
\end{equation}
where $\phi(\theta,\psi)$ is the normalized source brightness distribution function. The beam-weighted source solid angle is
\begin{equation}\label{omsigma}
\Omega_{\Sigma} \equiv \int_{source}  \phi(\theta,\psi)f(\theta,\psi)d\Omega
\end{equation}
where $f(\theta,\psi)$ denotes the normalized antenna power pattern \citep{Baars73}. Experiments have shown that we can approximate $f$ by gaussian beam. An exponential law of scale length $h=D_B/10$ is taken to model the source distribution function. The factor 10 is used to estimate the scale length of the molecular gas distribution from the optical diameter (B band). We used our own data from the CO(1-0) map of 3CR 31 and the CO(1-0) map of elliptical galaxies presented by Young (2002). In Eq.\ref{k} that optical diameter is represented by $\theta_s$.Usuing the asumption that the gas surface density is $\mu(r)\propto e^{-r/h}$ and as long as the source is smaller then the beam size, we have the following:
\begin{equation} \label{k}
K= \frac{{\int^{\theta_s/2}_{0}sin(\theta)e^{-^{10\theta}_{\theta_s}}d\theta}}{{\int^{\theta_s/2}_0 sin(\theta)e^{-^{10\theta}_{\theta_s}-ln(2)(^{2\theta}_{\theta_b})^2}d\theta}}
\end{equation}
If $I_{CO}$ is the velocity-integrated temperature for the $^{12}$CO(1-0) line  in $T_A^*$ scale and given the IRAM-30m parameters, the total mass of $H_2$ is then given by
\begin{equation}\label{mh2}
M_{H_2}=5.86 \times 10^4D^2KI_{CO} (M_{\odot})
\end{equation}
with the distance, D, in Mpc.\\
\begin{table*}
\caption{Molecular Gas data}             
\label{mol}      
\centering                          
	\begin{tabular}{l | lllllllll }        
			\textbf{Galaxy} &$Z_{CO}$&Distance& $I_{CO(1-0)}$ & $\delta I_{CO(1-0)}$ & FWZI$_{CO(1-0)}$ & $I_{CO(2-1)}$&$\delta I_{CO(2-1)}$& FWZI$_{CO(2-1)}$ &$M_{H_2}$\\
	     	&	& (Mpc) &(K Km/s) &(K Km/s)& (km/s) & (K Km/s)&(K Km/s)& (km/s)& ($ \times 10^8 M_{sun}$)	\\
	     	\hline
			$3C31^{**,a}$	& 0.0169& 71.06	& 5.68	& 0.81	&  550	& 14.06	& 1.13	& 550	& $16.81\pm2.38^{**}$	\\
			$3C40$					& ...	& ...	&$<$1.11&  ... 	&  ...	&  ... 	&  ... 	& ...	& $<3.25 $	\\
			$3C66B^{*,b}$	& 0.0157& 85.20	& 0.2081& 0.053 &  250	&  ... 	&  ... 	& ...	& $0.89 \pm 0.23^{*}$	\\
			$3C83.1^{*,c}$	& 0.0251& 104.6	&0.53	& 0.002 &  ---	&  1.13	& 0.12	& 200	& $3.60 \pm 0.0013$	\\
			$3C88^{**,b}$	& 0.03	& 126.20& 0.23	8& 0.05	&  300	&  ... 	&  ... 	& ...	& $2.19 \pm 0.44^{**}$	\\
			$3C98$					& ...	& ...	&$<$0.73&  ... 	&  ...	&  ... 	&  ... 	& ...	& $<7.27$	\\
			$3C129^{*,c}$	& 0.0208& 86.4	&0.28	& 0.024	&  ---	&  0.59	& 0.11	& 200	& $1.3 \pm 0.11$	\\
			$3C236$				& ...	& ...	&$<$0.75 &  ... &  ...	&   X 	&   X 	&  X	& $< 80.70$	\\
			$3C264^{**,a}$	& 0.02	& 90.90	& 0.70	& 0.15	&  200	& 1.48	& 0.16	& 225	& $3.37 \pm 0.74^{**}$	\\
			$3C270$				& ...	& ...	&$<$1.16&  ... 	&  ...	& ... 	&  ... 	& ...	& $<0.69$	\\
			$3C272.1^{*,a}$	& 0.0028& 12.00	& 0.36  & 0.12	& 200 	& 0.90	& 0.16	& 150	& $0.03 \pm 0.01^{*}$	\\
			$3C274^{**,b}$	& 0.0035& 14.90	& 4.02	& 0.39	&  200	&  ... 	&  ... 	& ...	& $0.52 \pm 0.05^{**}$	\\
			$3C296$				& ...	& ...	&$<$0.85&  ... 	&  ...	&  ...	& ... 	&  ... 	& $< 5.05$	\\
			$3C305^{**,b}$	& 0.042	& 171.70& 1.19  & 0.14	&  600	&   X 	&   X 	&  X	& $20.52 \pm 2.49^{**}$	\\
			$3C321^{**,b}$	& 0.10	& 379.97& 0.83	& 0.09	&  500	&   X 	&   X 	&  X	& $70.58 \pm 7.93^{**}$	\\
			$3C327^{*,b}$	& 0.1035& 401.87& 0.28	& 0.10	&  200	&   X 	&   X 	&  X	& $26.48 \pm 9.59^{*}$	\\
			$3C353^{**,b}$	& 0.0327& 153.73& 0.31	& 0.05	&  200	&  ... 	&  ... 	& ...	& $4.27 \pm 0.68^{**}$	\\
			$3C386^{**,c}$	&0.017	& 71.7	&0.58	& 0.02 	&  ---	& 1.23	& 0.17	& 175	& $1.78 \pm 0.058$	\\
			$3C402$				& ...	& ...	&$<$0.95&  ... 	&  ... 	&  ... 	&  ...	&  ...	& $< 6.59$	\\
			$3C403^{**,b}$	& 0.058 & 134.97& 0.44	& 0.11	&  500	&   X 	&   X 	&   X	& $4.75 \pm 1.18^{**}$	\\
			$3C433$				& ...	& ...	&$<$0.87&  ... 	&  ...	&   X 	&   X 	&   X	& $< 93.13$	\\
			$3C442^{**,c}$	& 0.0263& 110.5	& 0.09	& 0.02 	&  --- 	& 0.20	& 0.07	& 300	& $0.67 \pm 0.16^{**}$	\\
			$3C449^{**,b}$	& 0.0169& 71.40	& 1.20	& 0.24	&  500	& ... 	&  ... 	&  ...	& $3.59 \pm 0.73$	\\
			$3C465$				& ...	& ...	&$<$0.78&  ... 	&  ... 	&  ... 	& ... 	&  ...	& $< 7.69$	\\
			$NGC315^{*,b}$	& 0.0175& 73.80	& 0.26	& 0.06	&  150	&  ... 	&  ... 	&  ...	& $0.82 \pm 0.19^{*}$	\\
			$NGC326	^{*}$	& 0.0474& 203.18&0.2320&  0.05 	&  100	& 0.65 	&  0.2 	&  400	& $5.61 \pm 1.2^{*}$	\\
			$NGC541^{*,c}$	& 0.018	& 75.0	&0.72	& 0.08	& 400	& 1.53	& 0.32	&  350	& $2.5 \pm 0.27$	\\
			$NGC708^{**,b}$	& 0.0166& 69.86	& 1.84	& 0.31	&  600	& ...	&  ... 	&  ... 	& $5.25 \pm 0.88^{**}$	\\
			$NGC2484$				& ...	& ...	&$<$1.00&  ... 	&  ...	& ...	&  ... 	&  ... 	& $<19.77$	\\
			$NGC2892$				& ...	& ...	&$<$1.22&  ... 	&  ...	& ...	&  ... 	&  ... 	& $<6.85$	\\
			$NGC3801^{**,b}$& 0.0117& 49.50	& 3.82	& 1.14	&  600	& ...	&  ... 	&  ... 	& $5.49 \pm 1.63^{**}$	\\
			$NGC4278^{**,a^*}$& 0.0022& 9.40& 1.00	& 0.22	&  600	& 3.13	& 0.31	&  500	& $0.05 \pm 0.01^{**}$	\\
			$NGC5127^{**,c}$& 0.016	&67.1 	& 0.28	& 0.02	&  --- 	& 0.59	& 0.10	&  100	& $0.79 \pm 0.053$	\\
			$NGC5141$		& ...	& ...	&$<$1.55&  ... 	&  ... 	&  ... 	& ...	&  ...	& $<5.04$	\\
			$NGC5490$		& ...	& ...	&$<$1.92&  ... 	&  ... 	&  ... 	& ...	&  ...	& $<5.41$	\\
			$NGC6251$		& ...	& ...	&$<$1.03&  ... 	&  ...	&   X 	&   X 	&   X   & $<6.78$	\\
			$NGC7052^{**,a}$& 0.016	& 66.36	& 0.76	& 0.16	& 700   &  0.58	& 0.16	& 180   & $1.96 \pm 0.41^{**}$	\\
			$B2~0034+25$	& ...	& ...	&$<$0.96&  ... 	&  ... 	&  ...	&  ...	&  ... 	& $<10.41$	\\
			$B2~0116+31^{**,b}$	&0.06&242.10& 1.77	& 0.49	& 700	& 4.29	& 0.69	& 1000	& $60.63 \pm 16.92^{**}$\\
			$B2~0648+27^{**,a}$	&0.04& 169.83& 0.72& 0.07	& 300	& 1.44	& 0.34	& 350	& $12.25 \pm 1.18^{**}$	\\
			$B2~0836+29B^{**,b}$&0.065&261.70	&1.20&0.30	& 500	&  X 	&   X 	&  X	& $48.18 \pm 11.95^{**}$\\
			$B2~0915+32B$		& ... & ...	&$<$1.83&  ... 	&  ... 	&  ... 	& ...	& ...	& $< 70.52$	\\
			$B2~0924+30^{*,b}$	& 0.025& 104.25& 0.32& 0.11	&  200 	&  ... 	& ...	& ...	& $2.01 \pm 0.68^{*}$	\\
			$B2~1101+38$		& ...& ...	&$<$1.35&  ... 	&  ... 	&  ... 	& ...	& ...	& $<13.09$	\\
			$B2~1347+28^{*,b}$	& 0.072 & 297.9	&0.35&0.08 	&  ---	& 0.75	& 0.25	& 500	& $17.04^{*}$	\\
			$B2~1357+28$		& ...& ...	&$<$1.13&  ... 	&  ... 	&  ... 	& ...	& ...	& $<48.09$	\\
			$B2~1447+27$		& ...& ...	&$<$0.79&  ... 	&  ... 	&  ... 	& ...	& ...	& $<7.96$	\\
			$B2~1512+30$		& ...& ...	&$<$1.19&  ... 	&  ... 	&  ... 	& ...	& ...	& $<110.54$	\\
			$B2~1525+29$		& ...& ...	&$<$1.52&  ... 	&  ... 	&  ... 	& ...	& ...	& $<69.55$	\\
			$B2~1553+24$		& ...& ...	&$<$1.52&  ... 	&  ... 	&  ... 	& ...	& ...	& $<29.68$	\\
			$UGC7115$			& ...& ...	&$<$1.53&  ... 	&  ... 	&  ... 	& ...	& ...	& $<7.66$	\\
			$OQ208^{**,b}$		& 0.0766& 304.90&2.51&0.33	&  400	&  X 	&   X 	&  X	& $136.51 \pm 17.94^{**}$ \\
		\\
		 \end{tabular}\\
		The $^{*}$ means that this galaxy is a tentative detection, the $^{**}$ means this galaxy is detected. The letter \textit{a} is for (1-0)$\&$(2-1) detected galaxies, \textit{b} is for the CO(1-0) only detected galaxy and \textit{c} is for the galaxies detected in CO(2-1) only; note that in this cases the velocity with is represented as \textbf{---} since the CO(1-0) value was calculated using the average of the ratio and therefore the width of the line is unknown. The case of NGC4278, that has an $a^{*}$ is because this galaxy has been detected in (2-1) and tentatively detected in (1-0). In $M_{H_2}$ the $^*$ have the same meaning, only that in this case refers strictly to CO(1-0) since the integrated velocity for 3 mm was the one used to calculate the molecular mass.
\end{table*}
Using the survival analysis statistics software (ASURV), which takes into account the upper limits, besides the detections, we calculated a median value for the molecular gas mass of $2.2\times10^8~M_{\odot}$ (and an average of 3$\times 10^8$ M$_{\odot}$) for our complete sample of 52 galaxies. Their molecular gas mass distribution can be seen on Fig.\ref{mh2histo}. The molecular gas distribution is normalized to one for the whole sample including the detections and the upper-limits for CO(1-0) observations. On the same Figure we show in darker color the histogram of only the galaxies detected, the median value of the molecular gas mass of these 30 detected galaxies is then $3.4\times10^{8}~M_{\odot}$.\\ 
To normalize the sample we divided the distribution per bin of the histogram by the total number of galaxies in the sample, and to normalize the subsample, that includes the detected galaxies only, we divided the distribution per bin by the total number of galaxies in the complete sample, with the purpose of comparing both samples.  The range of the molecular gas mass (in Log) for the sub-group of the detected galaxies is between 6.3 and 10.2. The number of galaxies with a molecular gas mass lower than 6.3 (in Log) is obviously unknown.\\ 
The molecular gas mass in the sample ranges over 4 orders of magnitude, from  $3\times10^{6}~M_{\odot}$ for 3C272.1 to $1.4\times10^{10}~M_{\odot}$ for OQ 208. Most of the galaxies have a molecular gas mass  between 8.3 ($2\times10^8~M_{\odot}$) and 8.7 ($5\times10^8~M_{\odot}$), this is true for the complete sample including the upper limit calculation as for the subsample that includes only the detected galaxies.\\
If we look at the relation between the optical luminosity (L$_B$) and the molecular gas mass of the galaxies (M$_{H_2}$), as shown in Fig. \ref{fig:lbmh2}, we find that there is no correlation, which is interpreted as that the molecular gas mass is independent of the size of the galaxies. The FIR Luminosity (which is more detailed in section \ref{firco}), and the molecular gas mass, normalized by the optical luminosity has also no correlation, as can be confirmed on Fig. \ref{fig:lbmh2lfirlb}.
\subsection{CO(2-1)-to-CO(1-0) line ratios} \label{lineratios}
\begin{figure} 
\centering
\includegraphics[scale=0.45]{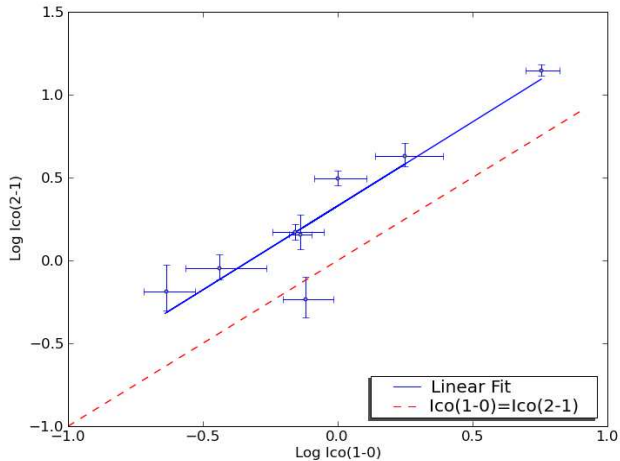}
\caption{Plot of the line ratio of the galaxies. The point that is outside of the linear fit is of the galaxy NGC 7052}\label{lr}
\end{figure}

%
The line ratio between the CO(2-1) and the CO(1-0) transitions is computed by comparing the integrated intensity ratio of the lines $I_{CO(2\rightarrow1)}/I_{CO(1\rightarrow0)}$ where the intensity was measured on one point at the center of each galaxy. 
Because of the different beam sizes at CO(1-0) and CO(2-1), the intensities should be compared only after having carried out small maps in CO(2-1) to sample the CO(1-0) beam. However, in most of our galaxies, the CO emission is expected to be almost a point source, at both frequencies. Except for NGC 4278, the distance of our detected galaxies is around or above 100 Mpc, where the CO(1-0) beam is 10 kpc, and in elliptical galaxies, the gaseous disks are expected to be, according to \cite{Young08}, concentrated in the nuclear disks at Kpc scales. Therefore, equal $I_{CO}$ intensities in both lines will result in ratios of $\sim$ 4, the point source dilution ratio of the two beams (to have a more detailed derivation of beam couplings, see Appendix \ref{coupling}).\\
As previously noticed by \cite{Lim2000}, the 2 galaxies studied in their paper have a stronger observed intensity in CO(2-1) than in CO(1-0). This sample has a line ratio well over unity. Seven of our galaxies have been detected in both frequencies and 6 more have been detected only in CO(2-1). From the galaxies detected in both frequencies, only NGC 7052 has an integrated intensity stronger in the CO(1-0) line than in the CO(2-1) line.  The maximum line ratio was found to be 3.1 (for NGC 4278) and the average value is 2.3 $\pm$ 0.1. When correcting by the factor 4 of the beam dilution ratio, the average value is $\sim$ 0.6. This corresponds to an average over a moderate density disk, where most of the CO emission is optically thick, but sub-thermally excited. The maximum ratio obtained precisely for the most nearby galaxy NGC 4278 is 0.8, comparable to the typical value obtained for the central parts of spiral galaxies by \citet{Braine92}.\\  
Figure \ref{lr} is a plot of the integrated intensity $I_{CO(2\rightarrow1)}$ vs. $I_{CO(1\rightarrow0)}$ with a line fit indicating a clear correlation. The dotted line in the plot is for the intensities with the same value for both lines. Table \ref{lrtable} lists the line ratios for each galaxy and to see the spectra of those galaxies detected in both frequencies see the online material, where the spectra of all galaxies in the sample are plotted. Note that the galaxy B2 0116+31 presents in its CO(1-0) line profile a strong absorption line, as well as the double line profile that can still be clearly seen. This absorption is the signature of molecular gas mass on the line of sight towards the AGN covering a very small area.  Most of the radio galaxies observed have a radio continuum in millimeter strong enough to be detected in absorption, but only in one galaxy the CO(1-0) transition is absorbed towards the radio continuum. The double horn profile of B2 0116+31 is visible as well in the CO(2-1) transition line, where the absorption is much weaker, as well as the continuum at this frequency. The absorption line in this galaxy should cause an underestimation of the CO(1-0) integrated intensity and therefore an overestimate of the line ratio.\\
\begin{table}
\centering
\begin{tabular}{l|c}
\hline
\hline
\textbf{Galaxy Name} & \textbf{Ratio}\\
\hline
\hline
3CR 31 & 2.5 $\pm$ 0.2\\
3CR 264 & 2.1 $\pm$ 0.2 \\
3CR 272.1 & 2.5 $ \pm$ 0.4 \\
NGC 326  &  2.8 $ \pm$ 0.3 \\
NGC 4278 & 3.1 $\pm$ 0.4 \\
NGC 7052 & 0.8 $\pm$ 0.1 \\
B2 0116+31 & 2.4 $\pm$ 0.3 \\
B2 0648+27 & 2.0 $\pm$ 0.3 \\
\hline
\end{tabular}
\caption{CO(2-1)-to CO(1-0) line ratio of the detected galaxies, not corrected by the different beam dilutions.}\label{lrtable}
\end{table}
For those galaxies detected in CO(2-1) emission line, and not detected in CO(1-0) we derived a value for the integrated velocity $I_{CO(1-0)}$ using the ratio of 2.3 calculated with the galaxies detected in both lines. Again, for a closer view to the spectra of those galaxies please see the online material.  The values of the derived molecular gas masses are already included on Table \ref{mol}.\\
%
\subsubsection{Origin of the low line ratio} \label{hlr}
%
As mentioned previously, the observations were done only in one point, at the center of the galaxies. However, the sources in the present sample are distant, and the expected extent of the CO emission is almost that of a point source, with respect to the large CO beams. \\
When we can appreciate from the HST optical images or from the CO interferometric map of 3CR 31 (see Fig. \ref{pdbi-3c31}), the  size of the molecular gas emission, it is possible to correct for the beam/source coupling ratio. \\
Taking into account the only 2 galaxies from this subsample for which we have the size of the molecular gas emission - 3CR 31 ($\S$ \ref{mgd}) and NGC 7052 ($\S$ \ref{ig}), we could roughly say that the size of the galaxies varies from 4'' to 8'', so that the correction factor for this subsample would be between 0.25 and 0.35. For 3CR 31 the result is more accurate since we have the CO map from the PdBI, and as can be derived from the map, we can see the galaxy has a molecular gas size of about 8'', implying a correction factor of about 0.3. If the line ratio of 3C R31 is 2.5, then after the correction, the ratio would be 0.8. This value is lower than one and therefore it could indicate that the CO line is sub-thermally excited \cite{Braine92}\\
In average, we find a value of $\sim$ 0.6 for the CO(2-1)-to-CO(1-0) line ratio.\\
This is compatible with a disk where the CO line is optically thick and sub-thermally excited, as is the disk of spiral galaxies in general. Only when it is possible to resolve the center of the disk, in nearby galaxies, the ratio has been observed to rise up to 0.8-1 \citep{Braine92}.\\

\subsection{Fanaroff and Riley classification} \label{frsection}

\begin{figure}
\centering
\includegraphics[scale=0.4]{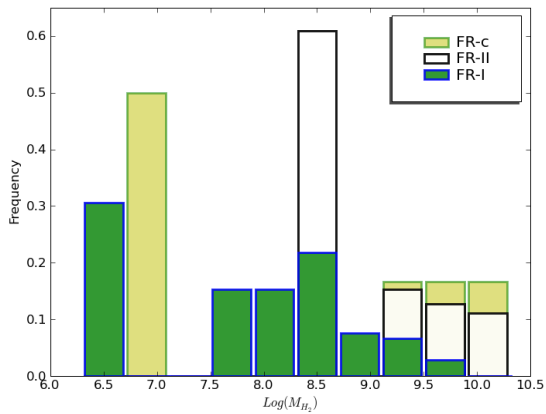}\\
\caption{Molecular gas mass distribution in the radio galaxies depending on their Fanaroff-Riley classification. 
}
\label{FRhisto}
\end{figure}

\begin{figure}
\centering
\includegraphics[scale=0.5]{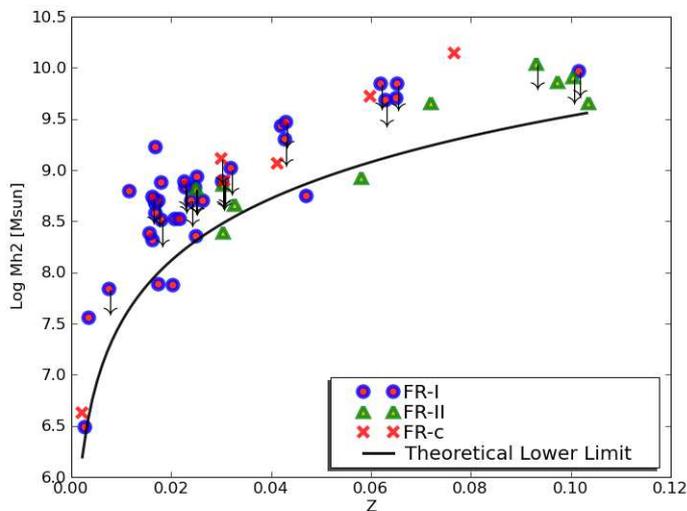}
\caption{$M_{H_2}$ versus z and a lower limit of the theoretical value for the mass with respect to the distance.}
\label{malmquist}
\end{figure}

Radio galaxies are divided mainly in two groups, the Fanaroff and Riley type I (FR-I) and type II (FR-II). As explained by \cite{Fanaroff74}, the sources were classified using the ratio of the distance between the regions of highest brightness of the radio continuum on opposite sides of the central galaxy or quasar, to the total extent of the source measured from the lowest contour; those sources from which the ratio was less than 0.5 were placed in class I, and those for which the ratio was greater than 0.5 were placed in class II. There is a third classification called FR-c galaxies. These galaxies have a very compact radio continuum emission.Their radio morphologies suggest that they are compact versions of the classical FR-II's, although why they are so small is not yet established: It is hypothesized that these are either young FR-II's or FR-II's trapped in a dense environment \citep{Fanti90,ODea91,Fanti94}.\\
In the present sample, 69.2\% are FR-I type galaxies; 19.2\% are FR-II type galaxies and 11.5\% are FR-c type galaxies.\\
We have previously said that the molecular gas mass average of our sample is $2.2\times10^8 ~ M_{\odot}$. FR-I and FR-II type galaxies do behave differently from one another. The FR-II type galaxies have the largest molecular gas mass compared to the FR-I type galaxies and the FR-c type galaxies, as shown in Table \ref{frmass}, using the detected galaxies and the upper limits with the survival analysis statistics. The mean molecular gas mass is lower for the FR-I type galaxies than for the FR-II type galaxies with $1.7\times10^8 ~ M_{\odot}$ and  $8.1\times10^8 ~ M_{\odot}$ respectively. This difference is clearly visible in Figure \ref{FRhisto} where the molecular gas mass distribution is shown according to the types of radio galaxies: FR-I, FR-II and FR-c.  We can draw the following conclusions regarding the different types of radio galaxies and their molecular gas masses:
\begin{enumerate}
\item  For the FR-I types, which are the less powerful AGN, the elliptical host galaxy does not need much molecular gas mass to host the radio AGN. Molecular gas masses can be as low as 10$^6$ M$_\odot$, which is a few Giant Molecular Clouds (GMC).
\item For the median value of the molecular gas mass, FR-II type galaxies ($8.1\times10^8~M_{\odot}$) are clearly more massive than the FR-I ($1.98\times10^8~M_{\odot}$) and FR-c ($2.0\times10^7~M_{\odot}$) type galaxies. Note that there are only a few FR-c galaxies and therefore the statistical values for them are not very reliable.
\end{enumerate}
\cite{Evans05} found only an 8\% detection rate of CO emission in their FR-II galaxies, a very low value compared to the $\sim$ 35\% detection rate of the FR-I detected in their sample. We also found a higher detection rate in FR-I galaxies compared to the FR-II galaxies. In our sample, the detection rate is 38.5\% for FR-I, 11.5\% for FR-II and 7.7\% for FR-c galaxies. Note that the FR-II being more powerful AGN, they are rarer and are found at larger distances. This can also explain both their lower detection rate, and their higher median molecular content.\\
To better estimate the Malmquist bias in our sample, we plot in Figure \ref{malmquist} the molecular gas mass vs. the redshift (z). It is clear that for higher z the galaxies tend to have higher molecular gas mass, following the sensitivity of the telescope. The sensitivity limit was computed, assuming a typical value of  300 km/s for the velocity width and 1 mK for the rms noise temperature. It is clear that there is a larger number of FR-II galaxies, compared to FR-I and FR-c galaxies, at higher z,  implying a larger threshold of the upper limit for the FR-II type galaxies. This also agrees with the idea that FR-II galaxies are stronger AGN and more luminous, and this is also why they are seen at a higher distance. A factor 3 in the median mass can be expected, and we conclude that the difference between the radio galaxy types for the molecular gas mass could be only due to the Malmquist bias. \\

\begin{table*}
\centering
\begin{tabular}{|c||c|c|c|c|}
\hline
&\multicolumn{2}{|c|}{\textbf{Detected $\&$ Upper Limits}}&\multicolumn{2}{|c|}{\textbf{Detected Only}}\\
\textbf{Galaxy Name} & \textbf{Mean} &\textbf{Median} &\textbf{Mean}&\textbf{Median}\\
\hline
\textbf{FR-I}	&$1.7\times 10^8~^{+2.3\times 10^8}_{-1.2\times 10^8}$	&${1.86\times 10^8}$	&$2.3\times 10^8~^{+3.2\times 10^8}_{-1.6\times 10^8}$	&${2.01\times 10^8}$\\
\textbf{FR-II}	&$8.1\times 10^8~^{+1.22\times 10^9}_{-5.3\times 10^8}$	&$4.45\times 10^8$		&$1.1\times 10^9~^{+1.7\times 10^9}_{-6.5\times 10^8}$	&$4.6\times 10^8$\\
\textbf{FR-c}	&$2.1\times 10^8~^{+9.2\times 10^8}_{-4.8\times 10^7}$	&$2.03\times 10^7$		&$8.5\times 10^8~^{+4.0\times 10^9}_{-1.8\times 10^8}$	&$1.2\times 10^{9}$\\
\hline
\textbf{Total}	&$2.3\times 10^8~^{+3.2\times 10^8}_{-1.68\times 10^8}$	&$2.2\times 10^8$		&$3.7\times 10^8~^{+5.2\times 10^8}_{-2.6\times 10^8}$	&$3.4\times 10^8$\\
\hline
\end{tabular}
\caption{Mean and median values of the molecular gas mass of all subsamples separately in logarithmic values in units of $M_{\odot}$}\label{frmass}
\end{table*}
%
\subsection{Molecular gas disk} \label{mgd}
The double horn CO line profiles observed are characteristic of an inclined molecular gas rotating disk. In the present set of data, the double horn feature is present in 8 galaxies: 3CR 31, 3CR 264, 3CR 403, 3CR 449, NGC 3801, NGC 7052, B2 0116+31 and OQ 208, i.e. 29\% of the detected galaxies in our sample. In the online material you can find the spectra of those galaxies that present a double horn profile. In the case of 3CR 31 and 3CR 264 the double horn profile is clearly visible in both transitions, CO(1-0) and CO(2-1), and for 3CR 403 and OQ 208 we see only this feature in the CO(1-0) transition since there is no data for the CO(2-1) transition.
Five of the galaxies that present the double horn profile are FR-I types. Only one galaxy, 3CR 403, is an FR-II type, and two galaxies, OQ 208 and B2 0116+31, are FR-c type.\\
We also note that 4 of our 8 galaxies have been detected in both transitions (3CR 31, 3CR 264, NGC 7052 and B2 0116+31) and in all 4 cases the CO(2-1) line is stronger than the CO(1-0) line (more details about the line ratios of these galaxies in $\S$ \ref{lineratios}).\\
The galaxies that present the double horn profile feature are also within the largest H$_2$ masses. The average value for this subsample is $29.1\times10^8~M_{\odot}$, an order of magnitude higher value than the average value of the complete sample for the detected galaxies only ($2.2\times10^8~M_{\odot}$).  \\
\begin{table}
	\centering
	\begin{tabular}{|l|c|}
		\hline
		\textbf{Galaxy} & \textbf{$M_{H_2}$}\\
		                &  ($\times10^8~M_\odot$)\\
		\hline
		3CR 31    & 16.81 $\pm$ 2.38 \\
		3CR 264   & 3.37  $\pm$ 0.74 \\
		3CR 403   & 4.75  $\pm$ 1.18 \\
		3CR 449   & 3.59  $\pm$ 0.73 \\
		NGC 3801  & 5.49  $\pm$ 1.63 \\
		NGC 7052  & 1.96  $\pm$ 0.41 \\
		B2 0116+31& 60.63 $\pm$ 16.92 \\
		OQ 208    & 136.51$\pm$ 17.94 \\
		\hline
		\textbf{Average Value:} & 29.14 $\pm$ 5.24 \\
		\hline
	\end{tabular}
	\caption{Molecular mass gas of the galaxies that present the double horn profile.} \label{mass-doublehorn}
\end{table}
We note that in some cases the absorption of the CO transitions  could mimic a double horn profile. It appears from the peak temperature in the double horn profiles that in all cases, except for the CO(2-1) emission in 3CR 31, the double horn profile could be perturbed by an absorption line.\\

\section{Dust} \label{dust}
\begin{figure} 
\centering
\includegraphics[scale=0.4]{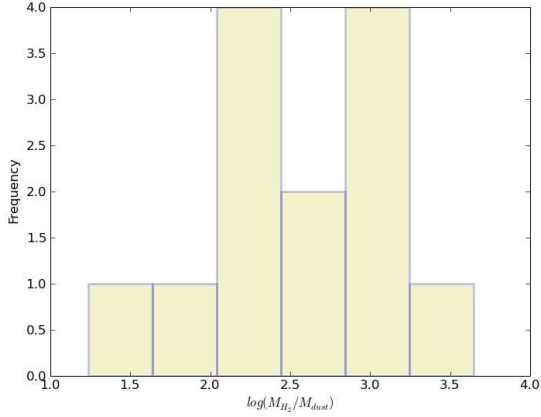}
\caption{Histogram of the Gas-to-Dust mass ratio.}
\label{gas-to-dust}
\end{figure}

\begin{figure*} 
\begin{tabular}{c c}
\centering
\includegraphics[scale=0.4]{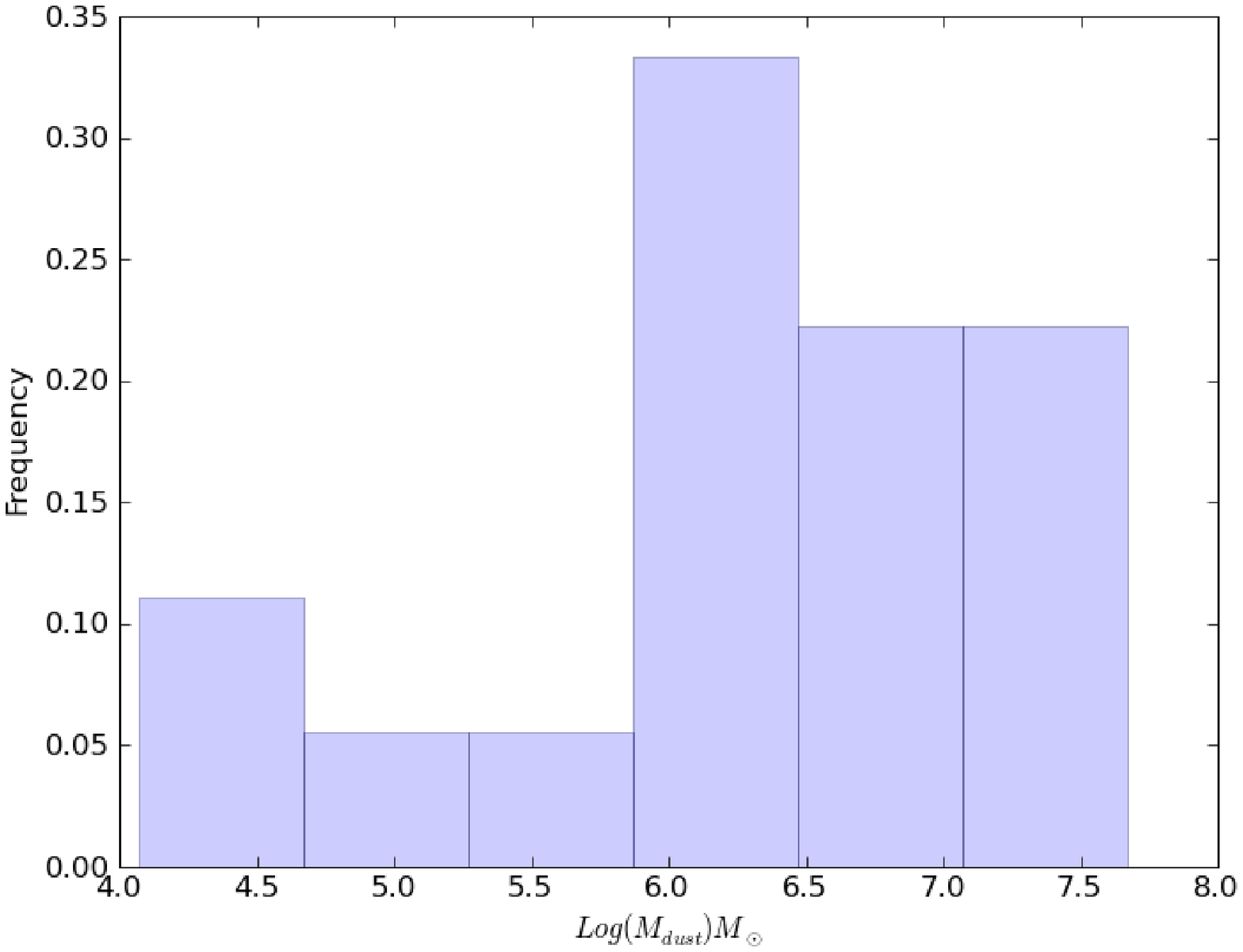}&\includegraphics[scale=0.4]{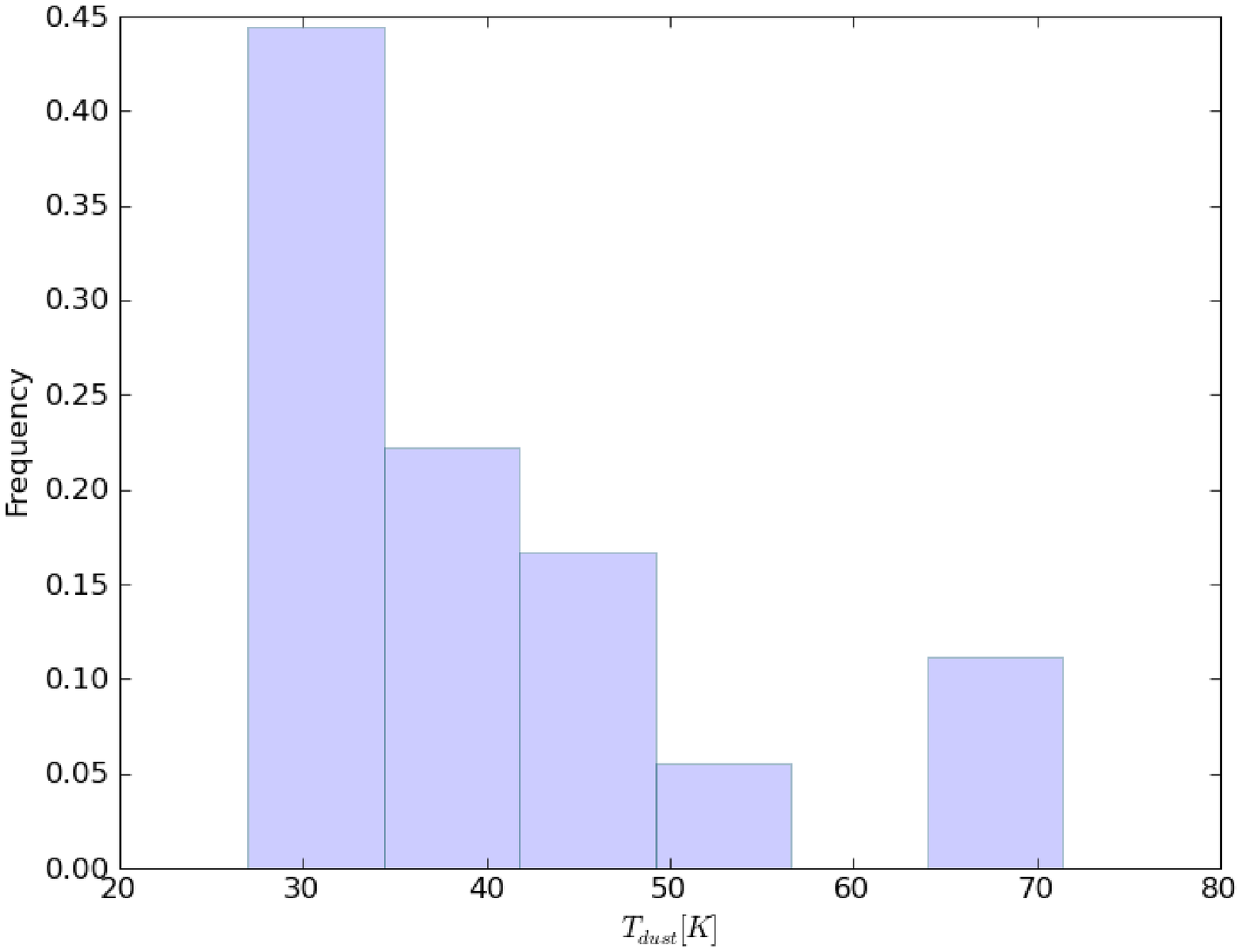}\\
\textbf{a)} & \textbf{b)}\\
\end{tabular}
\caption{Histograms of the dust characteristic of the sample. Figure a) is the dust mass in Logarithmic values and Figure b) is the dust temperature.} 
\label{DMTocana}
\end{figure*}

Dust is an important ingredient of the ISM, associated with the molecular gas. It is heated by  stars and/or an AGN, and its FIR emission depends strongly on its temperature, giving information about star formation (SF) in the galaxies. We analyze the dust emission at 60 $\mu$m and 100 $\mu$m in our sample. The fluxes at these frequencies give an estimation of the temperature and the mass of the warm dust component. On a companion paper we will use the fluxes at 12 $\mu$m and 25$\mu$m as well as the fluxes from the Spitzer satellite using both cameras, IRAC (3.6, 4.5, 5.8, 8 $\mu$) and MIPS (24,70, 160 $\mu$) to create the Spectral Energy Distribution (SEDs), and to study in more details the synchrotron and thermal emission in each galaxy of the sample.\\
The dust temperature $T_d$ is calculated by adopting a modified black-body emissivity law:
\begin{equation}
S_\nu \propto \kappa_\nu B_\nu(T_d)
\end{equation}
$\kappa_\nu$  is the dust emissivity and $B_\nu(T_d)$ is the Planck function at the dust temperature $T_d$. Then we calculate the dust mass as follows:
\begin{equation}
M_{dust}=4.78f_{100}D_L^2(e^{(143.88/T_d)}-1) ~ M_{\odot}
\end{equation}
Where $f_{100}$ is the flux at 100 $\mu$m  in Jy, $D_L$ the luminosity distance in Mpc and $T_d$ the dust temperature in K . The median value of the dust temperature for this sample, using the galaxies detected by IRAS, is 35.8K and the median value of the dust mass for the sample, only for the galaxies detected by IRAS, is $2.5\times10^6~M_{\odot}$. The detailed list of values for each detected galaxies at the frequencies of 60 $\mu$m and 100 $\mu$m is given in the  Table \ref{dust-data}.\\
The $M_{H_2}/M_{dust}$ gas-to-dust mass ratio for this sample, using the galaxies detected with both IRAS and IRAM-30m, is 260. On Figure \ref{gas-to-dust} the distribution of the gas-to-dust mass ratio is shown, where the galaxies range between 20 and 4450 in the gas-to-dust mass ratio. Comparing with the sample that includes the upper limits for the molecular gas mass gives us an upper limits for the ratio of 254 on average.
Figure \ref{DMTocana}a is the histogram of the dust mass of this sample. Note that most of the galaxies in this sample have a dust mass in the range $7.4-30\times10^5~M_{\odot}$ with 80\% of the detected galaxies having a dust mass greater than  $7.4\times10^5$. The dust temperature is mostly between 27.0 and 34.4 K (see Figure \ref{DMTocana} b) although there are 2 galaxies with an estimated dust temperature of 64 and 71.4 K respectively. These 2 galaxies with  a hotter dust component than the others are 3CR 327 and B2 0648+29B. These temperatures are derived assuming that the heated dust is mainly radiating by  the FIR traced by the IRAS fluxes at 60 and 100 $\mu m$ . Nevertheless we cannot discard a contribution of warmer and colder dust components than traced by IRAS. This issue will be addressed in a companion paper.
\begin{table}
\centering
\caption{Dust mass and temperature.} \label{dust-data}
\begin{tabular}{|l|c|c|}
\hline
\textbf{Galaxy Name}&\textbf{$Log(M_{dust})$} & \textbf{$T_{dust}$} \\
& ($M_{\odot}$) & (K) \\ \hline
 3C31 &  6.8 &28.3 \\
 3C88 & 7.1 &27.0 \\
 3C272.1 &4.6 & 36.4 \\
 3C274 (M87)& 4.1 & 47.4 \\
 3C305 &6.4 &39.5 \\
 3C321 & 7.1 &50.8 \\
 3C327 & 6.4 &71.3 \\
 3C402 &  7.0 &27.8 \\
 NGC 315 & 5.5 & 42.9 \\
 NGC 708 & 6.3 & 29.7 \\
 NGC 4278 & 5 & 20  \\
 NGC 6251 &  6.9 & 30\\
 NGC 7052 & 6.2 & 34.1\\
 B2 0116+31 & 7.4 & 29.2 \\
 B2 0648+27 & 6.2 & 68.5\\
 B2 0836+29B & 6.8 & 42.8\\
 B2 1101+38 & 6.2 & 35.3\\
 OQ 208 & 7.2 & 41.2\\
\hline
\end{tabular}
\end{table}
%
\section{FIR vs. CO} \label{firco}
%
\subsection{FIR $\&$ CO Luminosity}
%
We calculated the CO(1-0) line luminosity, $L'_{CO}$, which is expressed as velocity-integrated source brightness temperature, $T_b \Delta V$, times source area $\Omega_{s}D_A^2$, and which for our sample resulted in a mean value of $4.2 \times10^7~K~kms^{-1}pc^2$. This value was calculated using ASURV for the upper limits. The CO(1-0) luminosity mean value is 6.4 $\times10^7~K~kms^{-1}pc^2$ calculated only with  the detected values of the CO(1-0) emission lines. The formula used for the calculation is taken from \cite{Solomon97}:
\begin{equation}
L'_{CO}=23.5\times \Omega_{s\ast b}^2 D_L^2 I_{CO}(1+z)^{-3}
\end{equation}
Where $\Omega_{s\ast b}$ is the solid angle of the CO(1-0) emission for the source convolved with the telescope beam, $D_L$ the distance in Mpc, $I_{CO}$ is the integrated velocity in K km/s (in $T_{mb}$) and  z the redshift of the galaxy. If the source is smaller than the beam, which is the case for all the galaxies in this sample, we can assume that $\Omega_{s\ast b} \approx \Omega_b$ which is given in arcsec$^2$. \\
$L_{FIR}$ is the far-IR luminosity using the fluxes at 60 $\mu$m, associated with warm component, and 100 $\mu$m. associated with cooler components. It was computed using the following relation from \cite{Sanders96}:
\begin{equation} \label{Lfireq}
L_{FIR} = \biggl[ 1 + \frac{f_{100}}{2.58 f_{60}}\biggr]  L_{60}
\end{equation}
Where $f_{60}$ and $f_{100}$ are the fluxes at 60 and 100 $\mu$m respectively and are in units of Jy; $L_{60}$ is the luminosity at 60 $\mu$m in $L_{\odot}$ and it is represented as:
\begin{equation}\label{l60eq}
Log(L_{60}) = 6.014 + 2Log(D) + Log(f_{60}) 
\end{equation}
Where D is the distance of the galaxy in Mpc.\\
The results can be seen in Table \ref{lum} for which the median of the $L_{FIR}~=6.9\times10^9 ~ L_{\odot}$. This value was calculated using ASURV to take into account the upper limits.
\begin{table}
\centering 
\caption{Values of CO(1-0) Luminosities and FIR Luminosities} 
\label{lum}      
\begin{tabular}{l|l l}        
\hline
\textbf{Galaxy} & $Log(L'_{CO})$ &$Log(L_{FIR})$ \\
		& ($K~km~s^{-1}~pc^2$)	 & ($L_{\odot}$) \\
\hline
 3C31 & 8.51 & 9.77 \\
 3C40 & $<$7.86 & $<$9.30 \\
 3C66B & 7.23& $<$9.32 \\
 3C83.1 & 7.78 & $<$9.50 \\
 3C88 & 7.62 & 9.94 \\
 3C98 & $<$8.12 & $<$9.54 \\
 3C129 & 7.38 &  ...  \\
 3C264 & 7.81 & $<$9.46 \\
 3C270 & $<$7.13 &$<$ 8.51 \\
 3C272.1 & 5.77 & 8.23 \\
 3C274  & 7.01 & 8.25 \\
 3C296 & $<$7.99 &$<$ 9.53 \\
 3C305 & 8.60 & 10.21 \\
 3C321 & 9.12  & 11.43 \\
 3C236 & $<$9.10 &$<$ 10.28 \\
 3C327 & 8.69 & 11.24 \\
 3C353 &7.80   &  ...  \\
 3C386 & 7.52 &$<$ 8.75 \\
 3C402 & $<$8.07 & 9.90 \\
 3C403 & 8.43 & $<$10.62 \\
 3C433 & $<$9.17 &$<$ 11.20 \\
 3C442 & 7.08  &  ...  \\
 3C449 & 7.84 & $<$9.34 \\
 3C465 & $<$8.14 & $<$ 9.66 \\
 NGC 315 & 7.20 & 9.51 \\
 NGC 326 & 7.98 &$<$ 9.88 \\
 NGC 541 & 7.64 & $<$8.79 \\
 NGC 708 & 8.01 & 9.38 \\
 NGC 2484 & $<$ 8.54 &  ...  \\
 NGC 2892 & $<$8.10 &  ...  \\
 NGC 3801 & 8.03 & $<$9.60 \\
 NGC 4278 & 6.00 & 8.08 \\
 NGC 5127 & 7.14 &$<$ 8.45 \\
 NGC 5141 & $<$ 7.97 & $<$8.72 \\
 NGC 5490 & $<$ 8.00 & $<$8.74 \\
 NGC 6251 & $<$ 8.09 & 9.69 \\
 NGC 7052 & 7.58 & 9.65 \\
 B2 0034+25 & $<$8.27 &  ...  \\
 B2 0116+31 &  9.06 & 10.39 \\
 B2 0648+27 & 8.37 & 11.03 \\
 B2 0836+29B & 8.96 & 10.76 \\
 B2 0915+32B & $<$9.10 &$<$ 9.93 \\
 B2 0924+30 & 7.59 &  ...  \\
 B2 1101+38 & $<$ 8.37 & 9.75 \\
 B2 1347+28 &   8.50 &  ...  \\
 B3 1357+28 & $<$ 8.90 &  ...  \\
 B3 1447+27 & $<$ 8.16 &  ...  \\
 B3 1512+30 & $<$ 9.23 &  ...  \\
 B3 1525+29 & $<$ 9.06 &  ...  \\
 B3 1553+24 & $<$ 8.72 &  ...  \\
 UGC 7115 & $<$ 8.19 &  ...  \\
 OQ 208 & 9.41 & 11.12 \\
\end{tabular} 
\end{table}
%
\subsection{Star Formation}
The ratio of $L_{FIR}$/$L'_{CO}$ is normally related in spiral galaxies to the star formation efficiency (SFE).  This assumes that most of the FIR emission is coming from dust heated by young stars. In the case of powerful AGN, there is always the possibility of an added FIR contribution from dust heated by the AGN in a dusty torus. However, this contribution is rarely dominant \citep [see eg.][]{Genzel00}. Figure \ref{LfirLco_ocana} is a plot of the galaxies in our sample that represents the $L_{FIR}$ vs $L'_{CO}$ for the detected galaxies in both FIR and CO with their error bars, the galaxies with upper limits, represented with arrows (eg, down arrow is for the CO upper limit and FIR detection, right arrow is for the CO detected and upper limits for the FIR and the diagonal arrows are for the CO and FIR upper limits) the green line with the two circles are for when the $L_{FIR}$ was calculated with a 60$\mu$m flux detection and 100$\mu$m flux upper limit, representing the lower and upper limit for the $L_{FIR}$. \\
The linear relation between $L_{FIR}$ and $L'_{CO}$ is $log(L'_{CO})$ = $0.8 log(L_{FIR})+0.1$  with a $\chi^2$  of 0.024. Finally the relationship itself has a median value for the $L_{FIR}$/$L'_{CO}$  ratio of about 52 $L_{\odot}~(K~km~s^{-1}pc^2)^{-1}$. \\
We note that the detection rate for both FIR and CO together is independent of the type of radio galaxy. Out of the 15 radio galaxies detected for both FIR and CO, 8 (22\%) are FR-I (3CR 31, 3CR 272.1, 3CR 274, 3CR 305, NGC 315, NGC 708, NGC 7052 and B2 0836+29B), 3 (30\%) are FR-II (3CR 88, 3CR 321 and 3CR 327) and 4 (66\%) are FR-c (NGC 4278, B2 0116+31, B2 0648+27 and OQ 208).\\
We note as well that from the 15 galaxies detected for both FIR and CO, only 4 have a double horn profile feature (3CR 31, NGC 7052, B2 0116+31 and OQ 208). \\
As mention, $L_{FIR}/L'_{CO}$ is interpreted as a measure of SFE. Then, the Star Formation Rate (SFR) per unit gas mass, according to \cite{Gao04b}, is represented as:
\begin{equation}
\dot{M}_{SFR}\approx 2\times 10^{-10}~(L_{FIR}/L_{\odot})M_{\odot}yr^{-1}
\end{equation}
From this estimation we compute a mean SFR  for this sample of about  3.7 $\pm$ 1.5 $M_{\odot}/yr$, assuming that all the FIR emission comes from the dust heated by the young formed stars. This is likely  an overestimation since it does not take into account the AGN heating of the dust.\\
\begin{table*}
\centering 
\caption{FIR-to-CO Luminosities and their ratios.} 
\label{Lumfirco}      
\begin{tabular}{c|c c c }        
\hline
\textbf{Sample name} & $L_{FIR}$ &$L'_{CO}$ & $L_{FIR}/L'_{CO}$ ratio. \\
                     & ($L_{\odot})$ & ($K km/s pc^2$) &  ($L_{\odot}(K km/s pc^2)^{-1}$)\\
\hline
This sample & $8.9\times10^9$ & $4.2\times10^7$ & 52.0 \\
\cite{Wiklind95a} & $6.5\times10^{9}$ & $1.6\times10^{8}$ & 39.4 \\
\hline \\
\cite{Mazzarella93}& $1.94\times10^{11}$ & $1.1\times10^{9}$ & 82.2 \\
\cite{Evans05} & $3.6\times10^{10}$ & $1.2\times10^{9}$ & 35.4 \\
\cite{Bertram07}& $4.2\times10^{10}$ & $9.2\times10^{8}$ & 49.1 \\
\hline \\
\cite{Solomon97} & $1.2\times10^{12}$ & $7.9\times10^{9}$ & 146.8 \\
\hline
\end{tabular} 
\end{table*}
%
%
%
\subsection{CO-FIR relationship }
%
%
\begin{figure}
  \includegraphics[scale=0.45]{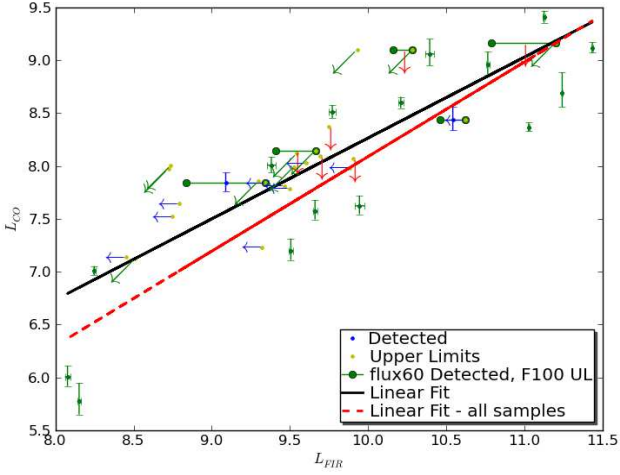}
  \caption{$L_{FIR}$ vs. $L'_{CO}$ for the galaxies in this sample. The upper limits are shown, where the L$_{CO}$ upper limits are represented by a down arrow, the L$_{FIR}$ upper limits are represented by a lef arrow and the L$_{CO}$-L$_{FIR}$ upper limits are represented by south weast arrows. It is also possible to appreciate the galaxies in which the flux at 100 $\mu$m was an upper limit and the flux at 60 $\mu$m was detected eith their error bars, in case L$_{CO}$ was detected and the down arrow in case L$_{CO}$ was upper limits} \label{LfirLco_ocana}
\end{figure}
\begin{figure}
	\includegraphics[scale=0.45]{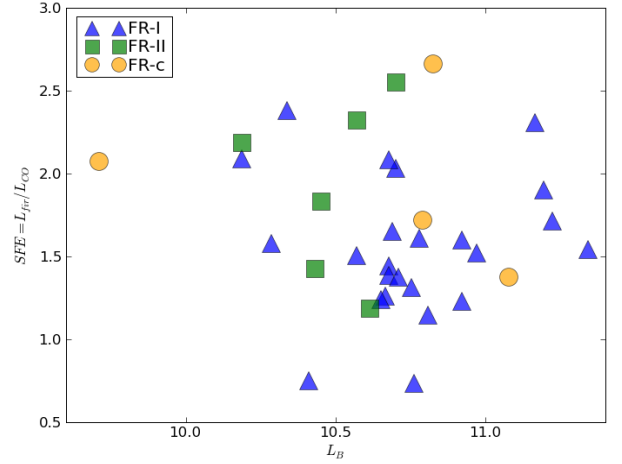}
	 \caption{Star formation efficiency compared to the blue luminosity of this sample of galaxies. } \label{figSfeLb}
\end{figure}

In systems with ongoing star formation (SF), the light from both newly formed and older stars can be absorbed by dust and reprocessed into the FIR. \cite{Bell03} states that since young stars in HII regions heat up dust at relatively high temperatures (with a low 100 $\mu$m-to-60 $\mu$m ratio of about $\sim$1) we could have an idea of where is this IR emission coming from, and this ratio leads to a wide range in$f_{100\mu m}/f_{60\mu m}$ on galaxy-wide scales, from $\sim$10 for early type spiral galaxies to $\lesssim$ 1 for the most intensely star forming galaxies suggesting that earlier types are influenced by old stellar population.\\
Our sample has a median value of $f_{100\mu m}/f_{60\mu m}{}={}1.9$ with its smaller value being 0.55 for 3C 327, which could be interpreted as galaxy with high temperature heated up by young stars in HII regions. And its highest value is 4.5 for 3CR 88, which is, still, a small value compared with $\sim$10 for quiescent early type spirals suggested by \cite{Bell03}.\\
Fig. \ref{LfirLco_ocana} shows a linear relation in our data for the $L'_{CO}$ and $L_{FIR}$. It is evident that the CO luminosities increase linearly with increasing $L_{FIR}$ with a relation such as $log(L'_{CO}){}={}0.8log(L_{FIR})-0.6$. On the plot we see there are two linear fits, the first one, the solid line, represents the linear fit of the galaxies from this sample only. On section \ref{sec:comp} we will compare our sample with other samples, this second fit, the dashed line, is the fit that we got from fitting all the galaxies together.\\
We have related the line ratio of the galaxies clearly detected in both lines with the gas-to-dust mass ratio as can be seen in the online section on Figure \ref{lr-k}. From this we see that most of the galaxies have a gas-to-dust mass ratio lower than the median value for the ratio of the complete sample. Recall that the average gas-to-dust mass ratio of the sample is $\sim$300.\\ The galaxy 3CR 88 has the lowest gas-to-dust mass ratio, 17, a very low value compared to NGC 4278 with a ratio of 600.\\
We also compared the SFE ($L'_{CO}$/$L_{FIR}$) vs. $L_B$, as shown in figure \ref{figSfeLb} where we found no correlation at all. In the figures the triangles are for the FR-I type galaxies, the squares are for the FR-II and the circles for the FR-c. The sample all together do not show correlations as well as all the subsamples separately. \\
%
%
\section{Comparison with other samples}\label{sec:comp}
\subsection{General description of the comparison samples}
We compare our sample with other samples selected under different criteria. 
\begin{itemize}
\item \cite{Evans05} observed  a sample of elliptical radio galaxies, like our own, but  selected with the IRAS fluxes densities at 60 $\mu$m, $f_{60 ~ \mu m}$, or 100 $\mu$m, $f_{100 ~ \mu m}$, greater than 0.3 Jy, since they assumed that this level is a good indicator of a dusty, gas-rich, interstellar medium.\\ 
\item \cite{Wiklind95a} observed a sample of elliptical galaxies selected on the basis of their morphological type. They argue that there is a major difficulty in defining a sample of elliptical galaxies due to uncertain morphological classification. According to them, the best criterion for defining an elliptical galaxy appears to be the Vaucouleurs $r^{1 \setminus 4}$ luminosity profile, and based on this, they selected genuine ellipticals, or galaxies that had a consistent classification as  \textit{E} in several catalogs. A 100 $\mu$m flux $\gtrsim$ 1 Jy and  $\delta  > 0^{\circ}$ were an additional criterion.\\
\item \cite{Mazzarella93} studied an infrared limited sample with 60 $\mu$m flux greater than 0.3 Jy. The objects were chosen to fully cover the sky, and to be relatively near (z $<$ 0.1), with no other selection criteria.\\
\item \cite{Bertram07} studied a sample of galaxies hosting low-luminosity quasi-stellar objects (QSOs) where they argue that an abundant supply of gas is necessary to fuel both, an active galactic nucleus and any circumnuclear starburst activity. The only criterion of selection was their small cosmological distance: only objects with a redshift z$<$0.060 were chosen.\\
\item  \cite{Solomon97} studied a sample of 37 infrared-luminous galaxies in the redshift range z=0.02-0.27. Eleven of these galaxies have a 60 $\mu$m flux $S_{60}>5.0$ Jy, and are part of the near bright galaxy sample \citep{Sanders88, Sanders91}. Twenty galaxies were chosen from a redshift survey \citep{Strauss92} of all IRAS sources with 60 $\mu$m fluxes $S_{60}>1.9$ Jy. And a few lower luminosity sources were also included in the sample.\\
\end{itemize}
\subsection{Molecular gas mass}
Comparing our sample with the other samples mentioned above, we noticed that \cite{Wiklind95a} and our sample have an average of the molecular gas mass of a few $10^8~M_{\odot}$ (using only detected galaxies, not the survival analysis median value) and that both samples are being hosted by elliptical galaxies; \cite{Mazzarella93}, \cite{Evans05} and  \cite{Bertram07} samples all have a median value of the molecular gas mass about $10^9~M_{\odot}$, their samples are FIR selected galaxies or galaxies in interaction, therefore with an expected  higher molecular gas mass than in the sample hosted by normal elliptical galaxies. They conclude that the majority of luminous, low-redshift QSOs have gas-rich host galaxies. We notice that from all galaxies in our sample the QSO, OQ 208, is the galaxy with the highest molecular gas mass ($1.4\times10^{10} ~M_\odot$). Finally the sample of \cite{Solomon97} has the highest average of molecular gas mass of $10^{10}~M_{\odot}$, with a sample of ULIRGs, where intense star formation is happening. Table \ref{tablecomp} gives a summary of the average molecular gas masses compared with other samples, and the characteristics of the samples.\\
Note that when we compare our sample with the others, we use the median value for the molecular gas mass that was calculated using only the detected galaxies for consistency with the other studies where  the upper limits were not used for their calculations.\\
If we compare now with a subsample of galaxies from TANGO itself, that have been detected in both, IRAS (for 60 $\mu m$ and 100 $\mu m$) and CO(1-0); we have a total of 15 galaxies with a median value of 1.22$\times10^9 M_{\odot}$. This is an order of magnitude higher than  with the median value of the molecular gas mass of the complete sample. It is in complete agreement with the statement that galaxies with higher FIR fluxes have more molecular gas mass than the galaxies dimmer in the FIR emission.\\
\begin{table*}
\centering
\begin{tabular}{|c|c|c|}
\hline
\textbf{Criteria of selection}&\textbf{Sample Name} & \textbf{Molecular gas mass} \\
			      & 		    & ($\times10^8$ M$_{H_2}$) \\
\hline
\hline
\textbf{Elliptical galaxies}&This sample & 3.7  \\
&\cite{Wiklind95a} & 6.8  \\
\hline
\textbf{FIR galaxies }&\cite{Evans05} & 37  \\ 
\textbf{and QSO's}&\cite{Mazzarella93} & 11  \\
&\cite{Bertram07} & 30  \\
\hline
\textbf{ULIRGS}&\cite{Solomon97} & 100  \\ 
\hline
\end{tabular}
\caption{Comparison samples and the median value of their molecular gas masses}\label{tablecomp}
\end{table*}
%
\subsection{Molecular gas disk}
The double-horned CO profile was previously noticed by other authors. We compare the fraction of double horn CO profile in our sample with the fraction in other samples. For instance, \cite{Evans05} noted a double horn CO emission line profile in many of the spectra. Out of nine detected galaxies, 5 present the double horn profile feature. They studied two cases for which they had interferometric data, finding that in one case this was caused by a CO absorption feature at the systemic velocity of the galaxy \citep{Evans99} and in the other case, 3CR 31, the molecular gas is distributed in a molecular gas disk \citep{Okuda05}. \cite{Wiklind95a} had one case of double horn profile out of 4 galaxies detected. \cite{Bertram07} found 11 double horn profile feature within their 26 detected galaxies and \cite{Mazzarella93} has 2 out of 4 galaxies detected with the double horn profile. In Table \ref{per} we give the percentage of molecular gas disk traced by the double horn profile for each sample. Like for the molecular gas properties, the closest sample to our radio galaxy sample is the one of \cite{Wiklind95a} composed by genuine elliptical galaxies. Thus it seems that the presence or not of a radio source is independent of the molecular gas properties, at the kpc scale, of the host elliptical galaxy. The FIR-selected samples have higher percentage of molecular gas disk than the sample selected only on the basis of the radio continuum properties. The dispersion value of the percentage of the molecular gas disks in each sample was computed using a Poissonian statistics. \\
We want to add that we do not discard the possibility of the double horn profile in these galaxies might be caused by the projection effect (i.e. inclination of the galaxy onto the sky plane).
\begin{table}
\centering
\begin{tabular}{l|c|c}
\hline
\textbf{Sample}&\textbf{Molecular gas disk}& \textbf{Dispersion}\\
               &        (\%)               &   (\%) \\ 
\hline
This sample & 27.6& 9.4 \\
\cite{Wiklind95a} & 25& 25\\
\cite{Mazzarella93} & 50& 35\\
\cite{Evans05} & 55.6&25\\
\cite{Bertram07} & 42.3&13\\
\hline
\end{tabular}
\caption{Percentage of the galaxies, within the detected galaxies of each sample of the double horn CO profile feature} \label{per}
\end{table}
\subsection{Dust}
\begin{figure*} 
\begin{tabular}{c c}
\centering
\includegraphics[scale=0.4]{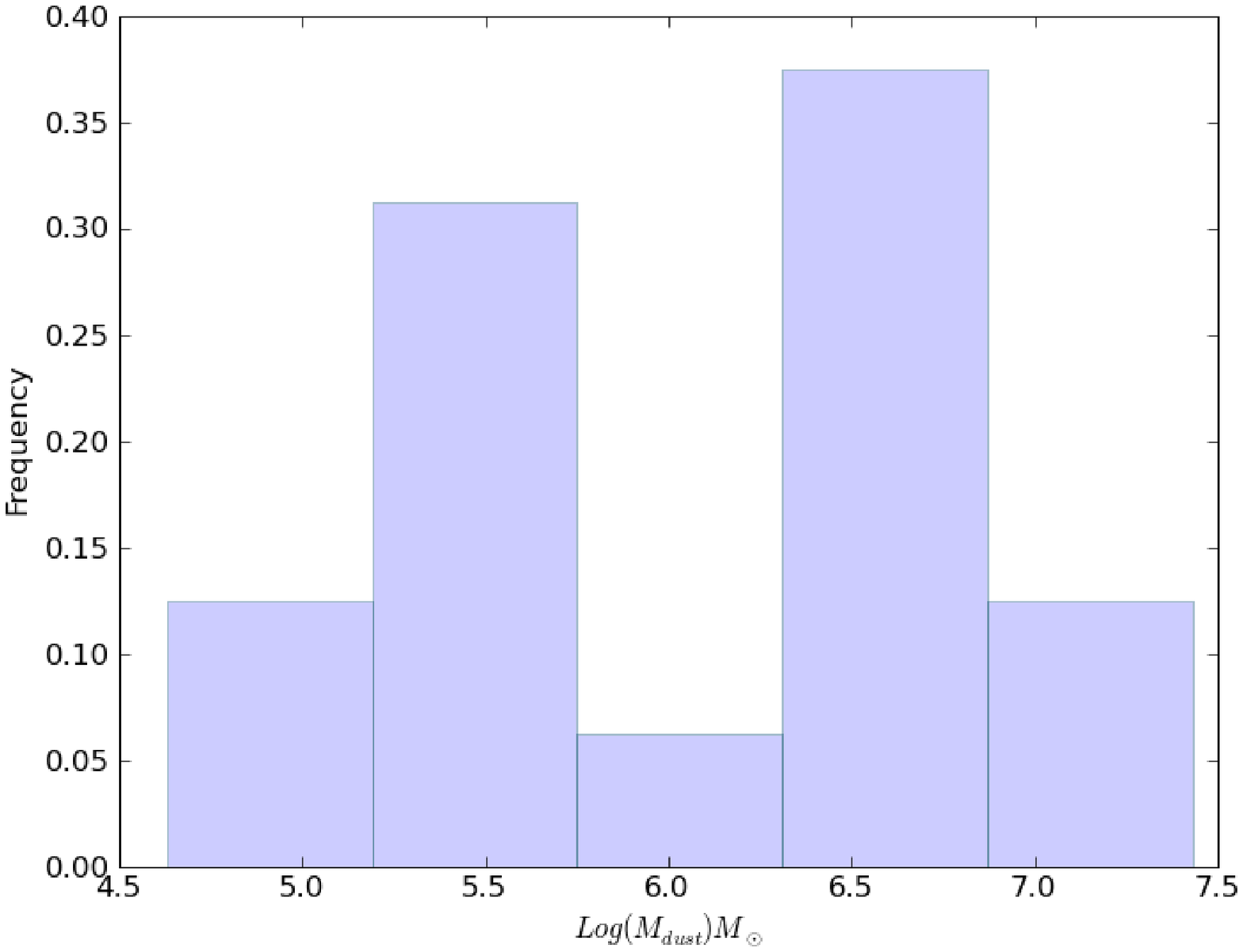}&\includegraphics[scale=0.4]{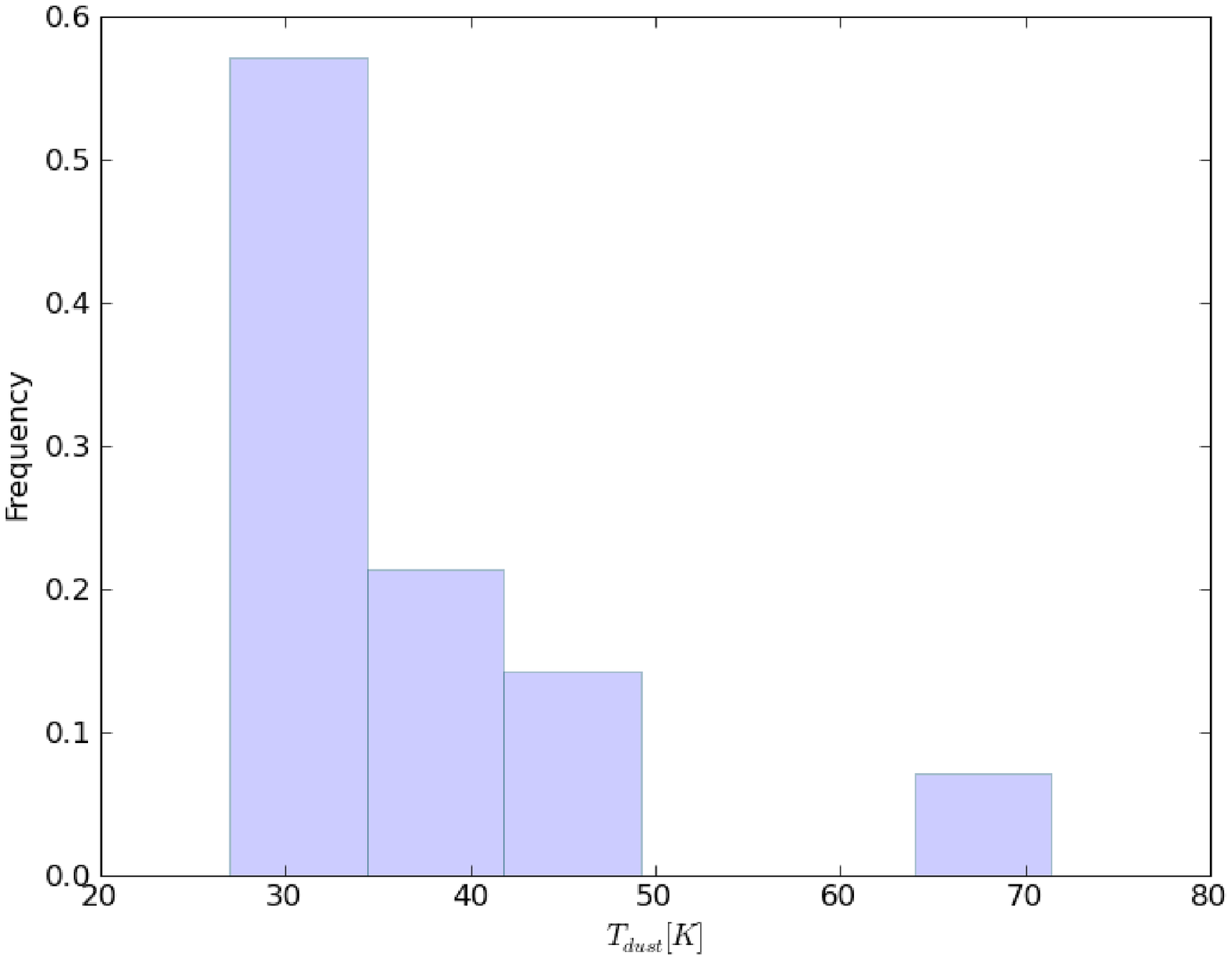}\\
\textbf{a)} & \textbf{b)}\\
\end{tabular}
\caption{Histograms of the dust mass and the dust temperature of the sample of \cite{Wiklind95a}. Fig. \textbf{a)} is the distribution of the logarithmic mass and Fig. \textbf{b)} is the distribution of the temeprature.}
\label{DMTwiklind}
\end{figure*}

We also compare the dust properties in our sample of radio galaxies, hosted by elliptical galaxies, with the sample of  elliptical galaxies from \cite{Wiklind95a} (see Fig. \ref{DMTocana} and \ref{DMTwiklind}). The dust component in our sample is slightly hotter than in the comparison sample of \cite{Wiklind95a} with a  median value of 34 K. The median value of the gas-to-dust mass ratio for \cite{Wiklind95a} is 700, much higher than the median value of the ratio in our sample, which is near 260. \cite{Evans05} found a gas-to-dust mass ratio even higher with a median value of 860.  The sample of low luminosity QSOs \citep[from][]{Bertram07} has a lower ratio of 452 and \cite{Mazzarella93} with their sample of radio galaxies detected by IRAS has a higher gas-to-dust mass ratio of 506.
\subsection{CO vs. FIR}
The sample of \cite{Solomon97}, composed of ULIRGs, is the sample with the highest $L_{FIR}/L'_{CO}$ of 146 $L_{\odot}(K km/s pc^2)^{-1}$ meaning, according to \cite{Solomon88}, that this group of galaxies are strongly interacting/merging galaxies with a strong SF, although this is mainly for spiral galaxies and we can not use this study to classify the rest of the galaxies here. Our sample has a ratio of 52, which is quite different from the ratio of 39 in the elliptical galaxies observed by  \cite{Wiklind95a}. \cite{Evans05} and \cite{Wiklind95a} seem to have similar ratios since \cite{Evans05} has a ratio of 35. Then \cite{Bertram07} have a ratio of 49.1 and finally \cite{Mazzarella93} with a very high ratio of 82.\\
\begin{figure}
\includegraphics[scale=0.5]{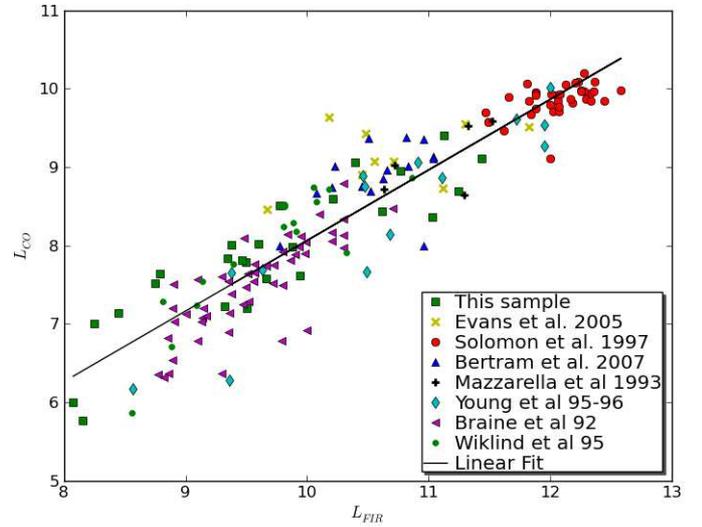}
 \caption{$L_{FIR}$ vs. $L'_{CO}$ of all the samples used in this paper. As can be appreciated from this plot, galaxies from \cite{Solomon97} which are ULIRGs, are the ones that better fit to the linear fit, as expected from the information in Table \ref{Lumfirco}.} \label{LfirLco_all}
\end{figure}
On Table \ref{Lumfirco} we give the median values of the CO(1-0) and FIR luminosity for the comparison samples together with the median value of the star formation efficiency (SFE=$L_{FIR}$/$L'_{CO}$).  As already known, \citep[see e.g.][]{Gao04b}, the ULIRG sample of  \cite{Solomon97} has the largest SFE indicator with a value of 147. With a median value of $L_{FIR}$ of $1.94\times 10^{11}$, the sample of radio galaxies of \cite{Mazzarella93} exhibits a high SFR tracing probably the star formation in merger galaxies. Similarly to the ULIRGs they show a quite large SFE indicator (82). The sample of low luminosity QSO host galaxies from \cite{Bertram07} as well as the galaxies in this sample, were chosen without any bias towards the FIR emission and their SFE appear to be similar with a value of 49 and 52 respectively. Finally the sample of elliptical galaxies \citep{Wiklind95b} and of FIR-selected radio galaxies \citep{Evans05} share a low value for the SFE indicator with 39 and 35.\\ 
The CO versus FIR luminosity is used in spiral galaxies to relate a tracer of the molecular gas, the CO emission, with a tracer of the star formation, the FIR emission. We compared the $L_{FIR}$ vs. $L'_{CO}$ relationship in  our sample with the one in  other samples. \\
A linear regression was fitted to the whole set of data (see Fig. \ref{LfirLco_all}) and to the individual samples. The linear regression is then log(L$_{CO}$)=m.log(L$_{FIR}$)+b; where m=0.9 and  b=-0.9 with a mean square error of 0.43. Considering the samples all together we obtain a $\chi^2$=0.022.\\
Since our data are well fitted by the CO-FIR relationship compared to the other samples, we can argue that the radio galaxies hosted by the elliptical galaxies  should exhibit at least a low level of star formation. \\
Besides the samples mentioned before, in Figure \ref{LfirLco_all} we include as well with 3 more samples: \cite{Young95} and \cite{Young96}, a sample of spiral galaxies observed as part of the FCRAO Extragalactic CO survey. We included also the sample of \cite{Braine92}, composed  of spiral galaxies. These samples were used to add well-defined samples for the CO-FIR relationship since the star formation is completely dominating that relationship in spiral galaxies.\\
%
\section{Discussion} \label{discussion}
%
\subsection{Origin of the molecular gas }
\begin{figure}
\includegraphics[scale=0.5]{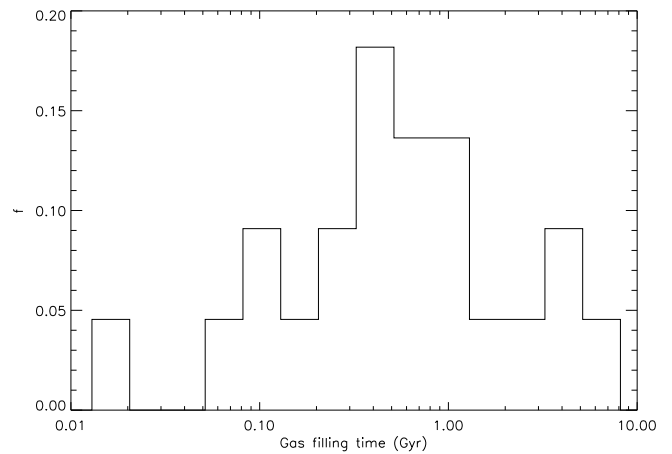}
 \caption{Histogram of the gas filling time for our sample. } \label{histo_fillingtime}
\end{figure}

\begin{figure}
\includegraphics[scale=0.5]{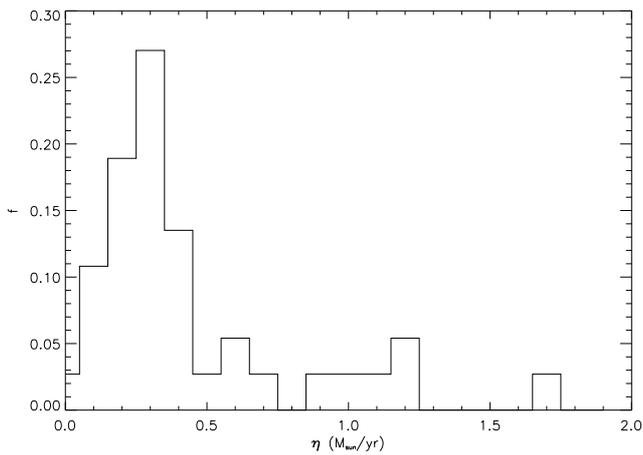}
 \caption{Histogram of the stellar mass loss in M$_\odot$/yr for our sample of radio galaxies. } \label{fig_stellarmassloss}
\end{figure}

%
As we have shown, the molecular gas content of the host elliptical galaxies of the powerful radio-AGN is low compared to spiral galaxies. The origin of that gas can either be internal (stellar mass loss, cooling gas from a galaxy halo of hot gas) or external (minor/major merger, accretion-fed from cosmological filaments, cooling flows), when the radio galaxy is at the center of a cluster. \cite{Salome08} provide an exhaustive discussion about the molecular gas in cooling flow which can be accreted by an elliptical galaxy. We estimate the stellar mass loss $\eta$ by using the B magnitude of the elliptical host \citep{Athey02}, namely $\eta = 0.0078(M_\odot /10^9 L_{\odot ,B}$) M$_{\odot}$/yr. Fig. \ref{fig_stellarmassloss}  shows the distribution of the stellar mass loss in our sample of radio galaxies detected in CO or with an upper limit of molecular gas mass lower than $5 \times 10^9 M_\odot$ in order to constrain the stellar mass loss contribution. The median value of the stellar mass loss is 0.37 $M_\odot. \mbox{yr}^{-1}$. This values implies a median time of 1 Gyr to refurbish the galaxy of molecular gas provided by the  stellar mass loss without taking into account the consumption rate of the AGN \citep[e.g.][]{David06}.  As shown on Fig. \ref{histo_fillingtime}, a substantial fraction of radio galaxies have a filling time $M(H_2)/\eta$ of molecular gas provided by the stellar mass loss larger than 1 Gyr (36\%) reaching time-scales close to the age of the Universe. It allows together with the accretion rate of the radio-AGN \citep[see e.g.][]{ Lagos09} to discard the stellar mass loss as a mechanism sufficient to provide the molecular gas in these elliptical galaxies. \\
In other words, we would expect the  gas lost by the stars to be hot and to enrich the hot halo gas through stellar wind, in spheroidal galaxies; spherical geometry dilutes the gas, such that the density is never high enough to cool. The cooling time begins $>$10Gyr, so must of this gas remains hot and will not be seeing in H$_2$/CO. This is not the case in spiral galaxies where the gravity and density are stronger so that the gas can cool back to the HI-H$_2$ phase.\\
If the molecular gas in the radio galaxies  was  provided by the stellar mass loss we would expect, at a first order, the molecular gas mass  to correlate with the blue luminosity,  a proxy of the galaxy stellar mass.That hypothesis is discarded by the data on Fig. \ref{fig:lbmh2} \\
The external hypothesis for the origin of the molecular gas is more likely to explain a large part of the molecular gas present in the elliptical galaxies hosting the radio-AGN. \cite{Lim2000} favored the minor merger scenario for the origin of the molecular gas in 3CR 31 because of the smooth isophotes which do no show any signs of recent perturbations caused by a major merger. However, 3CR 31 is the brightest galaxy in a cluster showing a cooling flow \citep{Crawford99}.\\
The inspection of the optical images shows that several radio galaxies are in dense environment where the minor/major mergers should be frequent enough to deposit molecular gas in the elliptical galaxies. The accretion of gas from cosmological filaments has been studied recently by numerical simulations \citep{Dekel09} and should be an additional source of molecular gas in these galaxies.\\
In the case of 3CR 31 there are suggestions, by \cite{Martel99} that the galaxy is interacting. \cite{Baldi08} suggest that 3C 264 has no young stellar population. \cite{Emonts06} propose that B2 0648+27 formed from a major merger event that happened $\lesssim$ 1.5 Gyr ago. \cite{GarciaBurillo07} state that the distortions in B2 0116+31 may reflect that the disk is still settling after a merger or an event of gas accretion; or maybe the jet and the cone-line features might be interacting with the disk producing the reported distortion. \cite{Giroletti05} states for NGC 4278 that its central BH is active and able to produce jets, however, the lifetime of its components of $<$100 yr at the present epoch, combined with the lack of large-scale emission, suggests that the jets are disrupted before they reach kilo-parsec scales. In conclusion, the origin of the gas in these galaxies can be multiple, in majority external, and a composite scenario can be invoked in each galaxy.\\

\section{Summary and Conclusion} \label{conclusion}
We present a survey of CO(1-0) and CO(2-1) lines for 52 radio galaxies in the Local Universe selected only on the basis of their radio continuum emission:
\begin{itemize}
\item The detection rate is 38\% (firm - 20/52) with 58\% total (30/52), including tentative detections. 
\item Indication of molecular gas disk (double-horn) is found in 15\% of the galaxies (27\% of the detected galaxies) with confirmation in 3CR31 with PdBI CO(1-0) map.
\item The CO(2-1)-to-CO(1-0) line ratio is 2.3 without taking into account beam dilution effect, and likely to be $\sim$ 0.6 with corrections, suggesting optically thick and sub-thermal CO emission.
\item The low ratio of $f_{100}/f_{60}$ also suggest that this sample may be forming stars.
\end{itemize}
We compare the properties of our sample with different samples of elliptical galaxies, QSO and ULIRGs:
\begin{itemize}
\item We find low molecular gas content compared to the other selected samples, with a median value of $2.2\times 10^8 ~ M_{\odot}$.
\item We confirm that FR-II type galaxies are characterized by lower CO detection rates than FR-I and FR-c radio sources as claimed by \cite{Evans05}, and that the FR-II have a higher molecular gas mass, as predicted from the existent Malmquist bias in this sample.
\item The $L_{FIR}$ vs. $L'_{CO}$ relation is similar to what is found in the other samples suggesting that some star formation could be taking place in these  elliptical radio galaxies.
\end{itemize}

\begin{acknowledgements}
	We thank Eric Emsellem  for a careful refereeing  and a  detailed report which helped to improve this paper significantly. B. Oca\~na Flaquer and  S. Leon  are partially supported by DGI Grant AYA2008-06181-C02-02, Spain. We thank Pamela Oca\~na for her careful readings and inputs on the manuscript.
\end{acknowledgements}
\bibliographystyle{aa}
\bibliography{ref2}
\Online

\begin{onecolumn}
{\LARGE\textbf{See Complete version with the spectra figures at \\ http://www.iram.es/IRAMES/personal/ocana/TanGoI\_v13.pdf }}
\end{onecolumn}

\begin{twocolumn}
\begin{appendix}

\section{Individual Galaxies} \label{ig}
 We present in this appendix various properties on the 29 detected galaxies, with increasing 3C numbers, then NGC and B2 numbers:
\begin{itemize}
\item \textbf{3C 31:} NGC 383, or Arp 331. For information on this galaxy, see $\S$ \ref{mgd}
\item \textbf{3C 66b:} UGC 1841. The well known bright radio jet of this galaxy was first discovered by \cite{Northover73} and the optical jet was first detected by \cite{Butcher80}. \cite{Meng06} were able to perform a detailed study of the jet using the HST/WFPC2 images comparing them with the high resolution radio images and finding excellent correspondence. The optical image reveals the presence of a disk, which is in agreement with the CO spectra of the galaxy. The  detection is a very faint one. The central disk diameter is about 10'' in the optical image, and the CO velocity width is 250 km/s.
\item \textbf{3C 83.1:} NGC 1265 (in Abell 426). It is one of the seven cases where the molecular gas has been clearly detected in the CO(2-1) transition and not in the CO(1-0) transition. The optical image shows a dust lane of  2.24'' length, according to \cite{Martel99}. This dust lane is oriented at $\sim$171$^{\circ}$, nearly orthogonal to the radio jet. If the CO emission corresponds to the dust lane, its small size explains the CO(1-0) signal dilution and non-detection.
\item \textbf{3C 88:} UGC 2748. It is clearly detected in the CO(1-0) transition. The optical image shows no peculiar feature.
\item \textbf{3C 129:} It is another example of CO(2-1) detection with an upper limit in the CO(1-0) line. The FWHM velocity width of this galaxy is 200 km/s. There is no hint of interaction in the Hubble image.
\item \textbf{3C 264:} NGC 3862 (in Abell 1367). This galaxy has a double horn profile, in both transition lines, typical of a molecular gas disk (see Figure \ref{onlineSpectra}). The velocity width of this galaxy is about 200 km/s. The disk profile can be seen in the optical image, which reveals also a jet \citep{Crane93}. \cite{Baum97} suggest that the optical synchrotron emission, clearly visible in the optical image from the HST, is associated with the jet.  According to \cite{Martel99} the nucleus is unresolved, and the host galaxy projected image is very circular and smooth. 
%
%
%
\item \textbf{3C 305:} IC 1065. In the CO(1-0) spectra, the detection is very clear, with a broad velocity width of $\sim$600 km/s. We have no data for the CO(2-1) emission line. The HST image presented by \cite{Martel99} shows filamentary and disturbed swaths of dust stretching across the western side of the galaxy and a twisting ``arm'' of emission in the eastern side. \cite{Jackson95} noticed a 1\arcsec scale cone-line extension of the continuum emission north of the core, in a direction almost 90$^{\circ}$ from the radio axis. They proposed that this might be an effect of obscuration of near-nuclear emission in all other directions rather than an example of scattering of anisotropic directed radiation from a hidden active nucleus.
\item \textbf{3C 321:} The HST optical image shows an extended dust absorption in the galaxy. It is clearly detected in the CO(1-0) transition with a velocity width of $\sim$500 km/s. For the CO(2-1) emission we have no data. The HST image reveals a nearby galaxy that might be interacting with this galaxy.
\item \textbf{3C 327:} It is clearly detected in the CO(1-0) transition line, with a FWHM of $\sim$200 km/s and with one of the highest dust temperature from this sample (64K).
\item \textbf{3C 353:} It is clearly detected in the CO(1-0) transition line, with a velocity width of $\sim$200 km/s.  \cite{Martel99} propose that the outer isophotes of 3C 353 are very circular while the inner isophotes are elongated in a southeast to northeast direction and are roughly peanut-shaped. \cite{Martel99} also stated that this may result from a small-scale dust lane bifurcating the nucleus in a rough north-south direction or from a true double nucleus. 
\item \textbf{3C 386:} It is clearly detected in CO(2-1) with a velocity width of $\sim$175 km/s, but not in CO(1-0). The HST optical image shows a bright optical nucleus where strong diffraction spikes dominate the core of the elliptical galaxy \citep{Martel99}.
\item \textbf{3C 403:} It is clearly detected in the only transition observed, CO(1-0), with a velocity width of $\sim$500 km/s. It presents the double horn profile already mentioned before and shown in Figure \ref{onlineSpectra}. \cite{Martel99} suggest this galaxy consists of two systems: a central elliptical region surrounded by a low-surface brightness halo with a sharp boundary at a distance of 3 Kpc northwest of the nucleus. In the northwest region of the halo, two or three very weak dust lanes are barely discernible. 
\item \textbf{3C 442:} UGC 11958, Arp 169. Detected in the CO(2-1) emission line, but not in the CO(1-0). The HST optical image shows the elliptical shape of the galaxy but no evidence of dust  obscuration. The isophotes of the outer halo of this elliptical galaxy are relatively smooth but within the central 520 pc, they are irregular and the light distribution becomes non uniform.
\item \textbf{3C 449:} It has a tentative double horn profile that can be noticed from the CO(1-0) spectrum (Figure \ref{onlineSpectra}), but one side is stronger than the other one. From the HST image of this galaxy it is possible to see the dust absorbing the visible light, not completely edge on. According to \cite{Martel99} the morphology of this galaxy suggests that we are viewing the near side of an inclined, geometrically thick torus or disk. The velocity width is about 500 km/s.
\item \textbf{NGC 315:} It is detected in CO(1-0) only, there is no visible feature in the HST image.
\item \textbf{NGC 326:} Detected with a higher intensity in CO(2-1) than in CO(1-0).
\item \textbf{NGC 541:} Arp 133 (in Abell 194). This galaxy has been detected in CO(2-1) but not in CO(1-0). From the HST image there is another galaxy that might be falling into NGC541 and there is another larger galaxy probably having some tidal effects on NGC 541. According to \cite{Noel03}, this cD S0 galaxy has a radio core on VLBA scales and a core-jet morphology on VLA scales. The central isophotes, measured from the WFPC/2 images,  vary considerably. The gas does not exhibit regular rotation profile.
\item \textbf{NGC 708:} in Abell 262. Detected in CO(1-0) but not in CO(2-1). This is the central galaxy of a cooling flow cluster, detected by \cite{Salome03}.
\item \textbf{NGC 3801:} It shows a double horn profile, detected in the CO(1-0) spectral line, with a velocity width of $\sim$600 km/s.
\item \textbf{NGC 4278:}  Also detected by \cite{Combes07}. The line emission is stronger in CO(2-1) than in CO(1-0), with a broader width. Note that this is the closest radio galaxy of all (9 Mpc). According to \cite{Combes07} the dust morphology across the disk consist in irregular patches or lanes; this suggests that this galaxy has recently accreted its gas \citep{Sarzi06}.
\item \textbf{NGC 5127:} There is a clear detection in the CO(2-1) line. The dust absorption is not clearly visible from the HST even though the galaxy is not so remote (z=0.016).
\item \textbf{NGC 7052:} It exhibits a double horn profile  very clear in the CO(1-0) emission (Fig. \ref{onlineSpectra}), but probably also in the CO(2-1) emission, although it is only a tentative detection there. 
According to \cite{Nieto90} the size of the dust disk is about 4\arcsec.
\item \textbf{B2 0116+31:} This is a peculiar galaxy in this sample, showing a strong absorption line in CO(1-0). The molecular/dusty disk has been studied in details by \cite{GarciaBurillo07} using the IRAM PdBI. The absorption line is surrounded by emission at both blue and red-shifted wings. Because of the absorption, we cannot discard a possible double horn profile. As mentioned in $\S$\ref{lineratios}, this absorption is the signature of molecular gas mass on the line of sight towards the AGN covering a very small area and because of its small filling factor, this absorption is quite rare, contrary to what could be expected for a radio-selected sample. Its double horn profile is visible as well in the CO(2-1) transition line, where the absorption is much weaker, as well as the continuum at this frequency, which was not detected; contrary to the CO(1-0) with a strong continuum flux of 164 mJy.. The absorption line in this galaxy should cause an underestimation of the CO(1-0) integrated intensity and therefore an overestimate of the line ratio.
\item \textbf{B2 0648+27:} It is detected in CO(1-0) \& CO(2-1) with a line ratio of 2.82 -the highest of all-. \cite{EmontPhd} Observed this galaxy with the VLA-C, where they detected HI in emission and in absorption noticing that it is distributed in a large and massive ring-like structure. This distribution and kinematic of the HI gas, together with the presence of faint tails in deep optical imaging \citep{Heisler94} and the detection of galaxy-scale post starburst young stellar population, imply that this galaxy formed from a major merger event that happened $\geq$1.5Gyr ago.
\item \textbf{B2 0836+29B:} It is clearly detected in the CO(1-0) emission line, but not in CO(2-1). Nothing peculiar in the optical image.
\item \textbf{B2 0924+30:} It shows a peculiar perturbed dust lane in the optical image which could possibly be due to interactions. It has not been detected in CO(2-1), although there is a clear detection of the CO(1-0) line. The HST image shows another smaller galaxy that might have fallen into B2 0924+30.
%
%
\item \textbf{B2 1347+28:} in Abell 1800. Detected in the CO(2-1) line but not in CO(1-0). The disk in the optical image does not seem to be perfectly elliptical, may be due to interactions.
\item \textbf{OQ 208:} Mrk 668. It has only been observed in the CO(1-0) line and it presents a clear double horn profile as shown in Figure \ref{onlineSpectra} with a velocity width of about 400 km/s. 
\end{itemize}
%
\section{Beam/source coupling} \label{coupling}

To correct the observed emission line brightness temperatures for beam dilution we know that (\cite{Thi04}):
\begin{equation}
T^{'}_{MB}=\frac{T_{MB}(\Omega_{source}+\Omega_{beam})}{\Omega_{source}}
\end{equation}
where $T^{'}_{MB}$ is the real main beam temperature and $T_{MB}$ is the observed main beam temperature. By assuming an axisymmetric source and beam distribution we can represent their distribution by an angular $\theta$ parameter leading to a correction factor for the CO(2-1)-to-CO(1-0) line ratio of:
\begin{equation}\label{klr}
K=\frac{ \theta_{b-230}^2+ \theta_s^2}{ \theta_{b-115}^2 + \theta_s^2}
\end{equation}
(see \cite{deRijcke06} for more details) where $\theta_{b-230}$ is the beam size at 230 GHz and $ \theta_{b-115}$ is the beam size at 115 GHz;  $\theta_s$ is the angular size of the source. We recall that for a gaussian beam  $\Omega_{beam}=\int_{beam}P_{\omega}\omega d\omega = \frac{1}{4 ln2} \pi \theta_b^2 \simeq 1.133 \theta_b^2$, although the 1.133 cancels out in Eq. \ref{klr}. The correction factor K for the CO(2-1)-to-CO(1-0) line ratio is shown on Fig. \ref{lr-k} as a function of the angular source size $\theta$, it varies between 0.25 for a point source up to 1. for a completely extended source.\\
Besides for the galaxy 3CR 31, we did not apply the beam dilution correction because  we dont have the molecular size of the galaxies and therefore we do not know which correction we should apply.\\
\begin{figure} 
\centering
\includegraphics[scale=0.4]{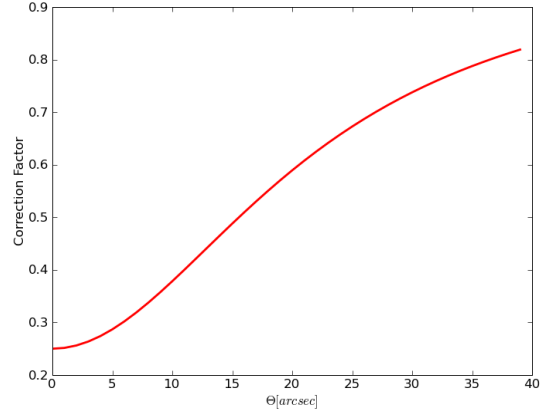}
\caption{Correction factor K for the CO(2-1)-to=-CO(1-0) line ration as a function of the size $\theta$ of the source detected in CO(1-0), to take into account the beam dilution. The possible sources go from a point like source to a source with a size of 22''}
\label{lr-k}
\end{figure}
\section{3CR 31} \label{3c31}
This galaxy has been observed in more details than the rest of the galaxies. It is an  FR-I radio galaxy at a distance of 71 Mpc hosted by an elliptical D galaxy in the Zwicky cluster 0107.5+3212. A double horn profile of the CO(1-0) and CO(2-1) lines was found by \cite{Lim2000} giving a first indication of a central molecular gas disk (see Fig. \ref{3c31-30m}). The Nobeyama Millimeter Array (NMA) CO(1-0) interferometer observations by \cite{Okuda05} confirmed the presence of this molecular gas disk. Our PdBI observations are about the same spatial resolution than the NMA ones but with a higher sensitivity as shown by the CO(1-0) integrated intensity map on Fig. \ref{pdbi-3c31}. The mean surface density of the molecular gas in the center is found to be $340 M_{\odot} . \mbox{pc}^{-2}$ given the spatial extension of 8\arcsec for the molecular gas disk as shown on the CO(1-0) intensity profile along the major axis (see Fig. \ref{3c31-Iprof}).\\
\begin{figure} 
\centering
\includegraphics[scale=0.30,angle=270]{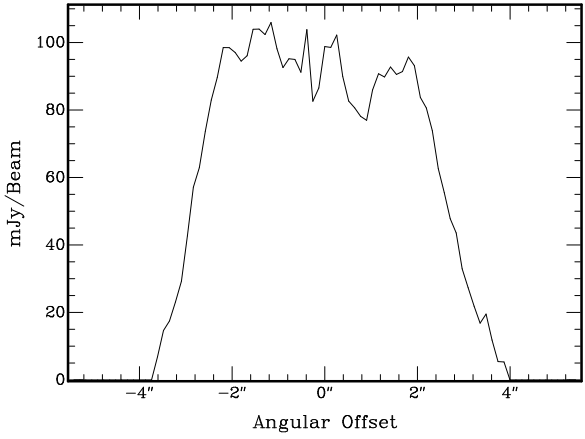}
\caption{CO(1-0) intensity profile along the major axis for the 3CR 31 galaxy.}
\label{3c31-Iprof}
\end{figure}
\begin{figure} 
\centering
\includegraphics[scale=0.40]{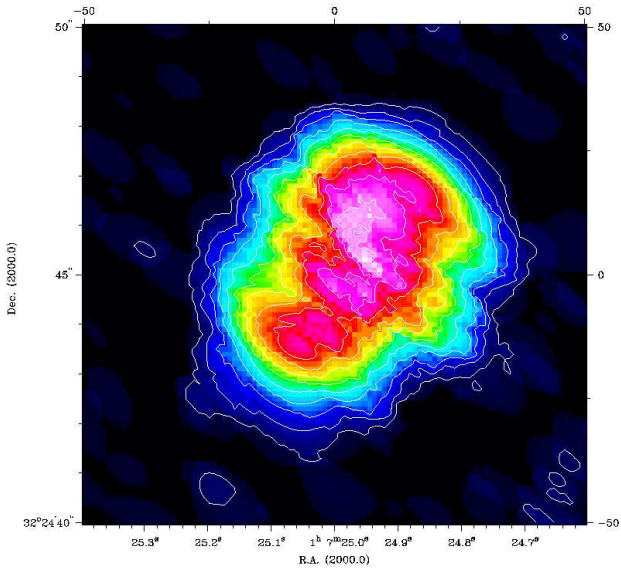}
\caption{CO(1-0) map of the PdBI for the  3CR 31 galaxy.}
\label{pdbi-3c31}
\end{figure}
\begin{figure} [h!]
\centering
\includegraphics[scale=0.4]{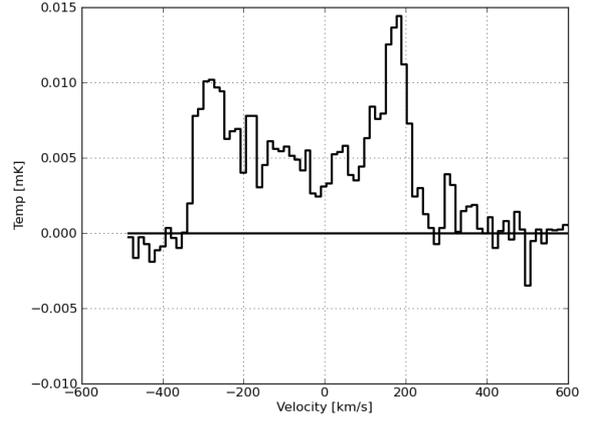}
\includegraphics[scale=0.4]{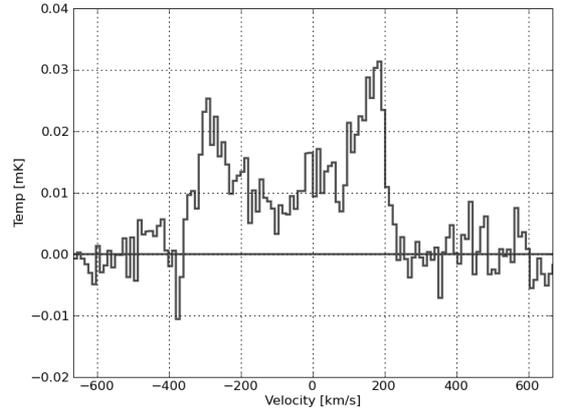}
\caption{3CR 31 spectra of the CO(1-0) (upper image) and CO(2-1) (lower image) lines.}
\label{3c31-30m}
\end{figure}
From the PdBI image (see Fig. \ref{3c31-Iprof} and \ref{pdbi-3c31}) we know that the molecular gas extension in 3CR 31 is about 8\arcsec  large. Applying the correction factor for the beam dilution (see Fig. \ref{klr}) to the CO(2-1)-to-CO(1-0) line ratio of 2.47  found with the IRAM-30m observations, the actual line ratio would go down to 0.8 assuming that the CO(2-1) is 8\arcsec  large. According to \cite{Braine92} a line ratio of about 0.7 implies an optically thick gas with an excitation temperature of about 7K.\\
\begin{figure} 
\centering
\includegraphics[scale=0.3,angle=270]{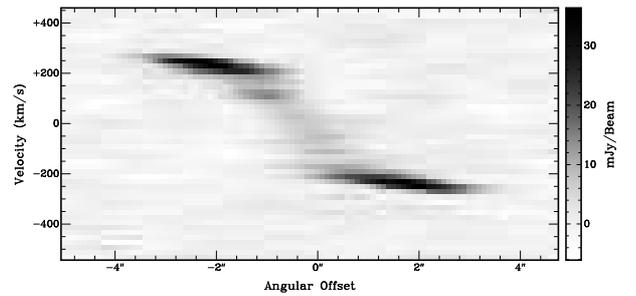}
\caption{PV-diagram along the major axis for the CO(1-0) emission in the 3CR 31 galaxy from the PdBI data. }
\label{3c31-pv}
\end{figure}
Figure \ref{3c31-pv} shows the position-velocity (PV) diagram for the CO(1-0) emission in 3CR 31. We can see that at 1\arcsec\ the maximum velocity $V_{max}$=190 km/s. Applying the correction for the inclination (i=39\degree) the maximum rotation velocity at a radius of 1\arcsec is $V_{rot_{max}}$=$V_{max}/\sin(i)~=$ 317 \kms.  Thus the dynamical mass inside a radius of 1\arcsec is estimated using $M_{dyn}=\frac{R*V_{rot}^2}{G}=1.02\times10^{10}~M_{\odot}$. We estimate the total molecular gas mass inside a radius of 1\arcsec\  to be $1.04 \times 10^8 M_{\odot}$ using the mean surface gas density, therefore the molecular gas represents  about 1\% of the  dynamical  mass in the very center of 3CR 31. \cite{Okuda05} discussed in detail about the dynamical and stability implications for the molecular gas in the center of 3CR 31.

\end{appendix}
\end{twocolumn}
\end{document}